\documentclass[12pt,draftclsnofoot,peerreview,onecolumn,notitlepage]{IEEEtran}
\IEEEoverridecommandlockouts

\usepackage{cite}
\usepackage{ieeefig}
\usepackage{amsfonts}
\usepackage{epsfig,color}
\usepackage{graphicx}
\usepackage[caption=false,font=footnotesize]{subfig}
\usepackage{stfloats}
\usepackage{amsmath}
\usepackage{paralist}
\usepackage[nolists,nomarkers]{endfloat}

\newtheorem{theorem}{\bf Theorem}
\newtheorem{lemma}{\bf Lemma}

\newtheorem{remark}{\bf Remark}

\newcommand{\define}    {\stackrel{\triangle}{=}}  

\begin{document}
%
\title{\huge{The Degrees of Freedom Regions of Two-User and Certain Three-User %
MIMO Broadcast Channels with Delayed CSIT} }

\author{Chinmay S. Vaze ~ and ~ Mahesh K. Varanasi
\thanks{
The authors are with the Department of Electrical, Computer, and
Energy Engineering, University of Colorado, Boulder, CO 80309-0425
USA (e-mail: {vaze, varanasi}@colorado.edu).

The material in this paper has been presented in part at the 2011 IEEE International Symposium on
Information Theory, St. Petersburg, Russia.} }

%



\maketitle
\pagestyle{empty}

\begin{abstract}
The degrees of freedom (DoF) region of the fast-fading MIMO (multiple-input multiple-output) Gaussian broadcast channel (BC) is studied when there is delayed channel state information at the transmitter (CSIT). In this setting, the channel matrices are assumed to vary independently across time and the transmitter is assumed to know the channel matrices with some arbitrary finite delay. An outer-bound to the DoF region of the general $K$-user MIMO BC (with an arbitrary number of antennas at each terminal) is derived. This outer-bound is then shown to be tight for two classes of MIMO BCs, namely, (a) the two-user MIMO BC with arbitrary number of antennas at all terminals, and (b) for certain three-user MIMO BCs where all three receivers have an equal number of antennas and the transmitter has no more than twice the number of antennas present at each receivers. The achievability results are obtained by developing an interference alignment scheme that optimally accounts for multiple, and possibly distinct, number of antennas at the receivers.
\end{abstract}
%




\section{Introduction}

\IEEEPARstart{T}{he} capacity region of the MIMO BC was obtained in \cite{Shamai-W-S} under the assumption of perfect (and instantaneous) CSIT. Under this idealized assumption, the degrees of freedom (DoF) region -- defined as the set of high ${\rm SNR}$ slopes of the rate-tuples in the capacity region relative to $\log({\rm SNR})$ -- which denotes the set of highest, simultaneously accessible fractions of signaling dimensions by the users, for a MIMO BC with $M$ transmit antennas and $N_i$ receiver antennas at receiver $i$ is characterized by the maximum sum-DoF being  $\min \{ M, \sum_{i=1}^K N_i \}$. On the other hand, without any CSIT whatsoever, and for a broad class of fading channel distributions, the DoF region collapses to what can be achieved just through time-division \cite{Vaze_Dof_final} (see also \cite{Jafar-Goldsmith,Chiachi2}), so that the sum DoF collapse to $\min \{ M, \max_i N_i \}$.

Moreover, even for the practically realizable assumption of imperfect CSIT, assumed fixed relative to ${\rm SNR}$, the maximum sum-DoF known to be achievable in the most general case does not improve the situation over that achievable without CSIT. One approach for gaining insight about the DoF behavior under imperfect CSIT has thus been to seek higher DoF under {\em quantized} CSIT by improving CSIT quality through a sufficiently fast scaling of the quantization rate with ${\rm SNR}$ \cite{Jindal,Ravindran,Caire-Jindal,Vaze_fb_scaling_GBC}. Another recent approach that is well-suited for channels, with high user mobility for example, in which the coherence time is relatively short compared to the delay incurred in channel estimation and feedback, is that of DoF characterization under delayed CSIT proposed by Maddah-Ali and Tse in \cite{maddah_ali_tse_delayed_CSIT}. The authors of \cite{maddah_ali_tse_delayed_CSIT} prove a surprising and interesting result that even in an i.i.d. fading channel -- in which predicting the current channel state based on past channel states would be meaningless -- if the transmitter has delayed (hence `stale') but perfect CSI, significant gains are possible in the achievable rates relative to the situation of complete lack of CSIT to such an extent that even the DoF are strictly higher. For example, the Gaussian BC with $2$ transmit antennas and two single-antenna users is shown to have $\frac{4}{3}$ sum-DoF compared to $1$ sum-DoF without CSIT.

The main idea in the achievability scheme of \cite{maddah_ali_tse_delayed_CSIT} is that the interference experienced by a user at a previous time, which is a linear combination of data symbols intended for some other user, can be evaluated perfectly by the transmitter at the current time using delayed CSIT and subsequently transmitted to provide the interfered user the opportunity to now subtract that interference while simultaneously sending a new linear combination of the data symbols to the user where these symbols are desired. Moreover, the schemes of \cite{maddah_ali_tse_delayed_CSIT} based on this principle were also shown to be sum-DoF optimal, i.e., the achievable sum-DoF is as high as an upper bound on the sum-DoF derived therein in the case where the number of transmit antennas is greater than or equal to the number of users.

In this paper, the results of \cite{maddah_ali_tse_delayed_CSIT} obtained for the MISO BC (i.e, the BC with single antenna receivers) are extended to the MIMO BC with an arbitrary number of antennas at each terminal. In particular, an outer-bound to the DoF region of this general $K$-user MIMO BC with delayed CSIT is obtained. For this outer-bound, two alternate proofs are provided. Like the proof in \cite{maddah_ali_tse_delayed_CSIT}, our first proof is based on (a) the result that feedback doesn't improve the capacity region of the degraded BC \cite{Gamal_fb_capacity_degraded_BC} and (b) the DoF region of the general $K$-user MIMO BC obtained by the authors under the assumption of no CSIT (under independent and isotropic distribution of channel directions) in \cite{Vaze_Dof_final}. Our second proof uses generic techniques of information theory (as opposed to the specialized result of \cite{Gamal_fb_capacity_degraded_BC}) and hence can be seen as being potentially more widely applicable. For example, the counterpart of this latter proof for the two-user MIMO interference channel (IC) with an arbitrary number of antennas at each of the four terminals under the delayed CSIT assumption results in a tight outer bound on the DoF region of that network and was recently found by the authors in \cite{Vaze_Varanasi_delay_MIMO_IC}. Furthermore, the outer-bound for the $K$-user MIMO BC is shown to be tight for the two-user MIMO BC by specifying a DoF-region-optimal achievability scheme based on interference alignment. Since there may be an unequal numbers of antennas at the receivers, the DoF {\em region} metric is appropriate rather than sum-DoF (which is sufficient when receivers have an equal number of antennas). The key idea behind the achievable scheme for the two-user MIMO BC is that when the transmitter sends interference caused at one or both receivers, the transmit signal must be constructed to account for the number of antennas at various terminals in a manner so as to cause no additional interference to any of the users while delivering the maximum number of linear combinations of data symbols to the receiver where those symbols are desired.

Moving beyond the two-user MIMO BC, we also study a special class of three-user MIMO BCs in which all three receivers have the same number of antennas and the transmitter has no more than twice the number of antennas at each receiver. For this class of MIMO BCs, \cite{abdoli_3user_BC_delayed_isit} recently determined the sum-DoF with delayed CSIT. Leveraging the scheme of \cite{abdoli_3user_BC_delayed_isit} and DoF region optimal scheme for the general two-user MIMO BC, we expand the sum-DoF result of \cite{abdoli_3user_BC_delayed_isit} to establish the exact DoF {\em region} of this special class of three-user MIMO BCs. For this class of MIMO BCs, the outer bound obtained here is also tight.

In addition to the complete DoF region of the 2-user MIMO IC in \cite{Vaze_Varanasi_delay_MIMO_IC}, there are several recent works that provide interference alignment schemes for networks with distributed transmitters under the delayed CSIT assumption
\cite{Jafar_Shamai_retrospective_IA, Ghasemi_Motahari_Khandani_MIMO_Delayed, Ghasemi_Motahari_Khandani_X_Delayed}. However, no tight outer bounds are provided and hence, by themselves, those results are inconclusive. Finally, we mention that the connection between the achievable schemes of \cite{maddah_ali_tse_delayed_CSIT} and blind interference alignment previously obtained in \cite{Jafar_correlations,Jafar_Gou_blind_IA_2010} are explored in \cite{Jafar_Shamai_retrospective_IA} which was also the first paper to provide examples of (retrospective) interference alignment schemes for some networks with distributed transmitters.

The next section describes the channel model and states the main results. The ensuing sections contain the proofs of these results, while the final section concludes the paper.

\section{Channel Model and Main Results}
In this section, we describe the model of MIMO BC under delayed CSIT and state our main results.

\subsection{The MIMO Gaussian BC}
Consider the MIMO Gaussian BC with $M$ transmit antennas and $K$ users having $N_1$, $N_2$, $\cdots$, $N_K$ receive antennas, respectively. Without loss of generality, it is assumed that $N_1 \geq N_2 \geq \cdots \geq N_K > 0$. The input-output relationship is given by
\begin{equation}
Y_i(t) = H_i(t) X(t) + Z_i(t),
\end{equation}
where at the $t^{th}$ channel use, $X(t) \in \mathbb{C}^{M \times 1}$ is the transmit signal, $Y_i(t) \in \mathbb{C}^{N_i \times 1}$ is the signal received at the $i^{th}$ user, $H_i(t) \in \mathbb{C}^{N_i \times M}$ is the corresponding channel matrix, and $Z_i(t) \sim \mathcal{C}\mathcal{N}(0,I_{N_i})$ is the complex additive white Gaussian noise. The transmit power constraint is taken to be $\mathbb{E} ||X(t)||^2 \leq P, \; \forall \, t$. Further, it is assumed that the channel matrices $H_i(t)$ undergo independent and identically distributed (i.i.d.) Rayleigh fading, i.e., the channel matrices are i.i.d. across $t$ and $i$, and their entries are i.i.d. standard complex normal $\mathcal{C}\mathcal{N}(0,1)$ random variables.

It is assumed that every receiver has perfect CSI (i.e., the knowledge of all channel realizations) and the transmitters have perfect CSI but with some finite but otherwise arbitrary delay which, without loss of generality, can be taken to be one time unit. In particular, the channel matrices $H_i(t)$ for every $i$ are known to all transmitters at time $t+1$. We refer to this assumption about channel state knowledge as {\em delayed CSIT}.

Consider any coding scheme that achieves the rate tuple $(R_1,R_2,\cdots,R_K)$. Let ${\cal M}_i$ be the message to be sent to user $i$ over a blocklength of $n$. We assume that the messages are independent and message ${\cal M}_i$ is distributed uniformly over a set of cardinality $2^{nR_i}$. We say that the rate tuple $(R_1,R_2,\cdots,R_K)$ is achievable if, at every user, the probability of error in decoding the respective message goes to zero as the blocklength $n \to \infty$. The capacity region $\mathcal{C}(P)$ is then defined to be the set of all achievable rate tuples $(R_1,R_2,\cdots,R_K)$ when the transmit-power constraint is $P$, while the DoF region is defined as follows:
\begin{eqnarray*}
\mathbf{D}^{\rm{d-CSI}} = \left\{ \Big( d_i \Big)_{i=1}^K \left| \forall ~i, d_i \geq 0 \mbox{ and }  \exists ~  \Big( R_1(P), \cdots,  R_n(P) \Big) \in \mathcal{C}(P) \mbox{ such that } d_i = \mathrm{MG} \Big( R_i(P) \Big) \right. \right\},
\end{eqnarray*}
where the function `multiplexing gain' $\rm{MG} (\cdot)$ is defined as $\rm{MG}(x) = \lim_{P \to \infty} \frac{x}{\log P}$.

\subsection{Main Results}
The following theorem gives an outer-bound to the DoF region of the MIMO BC with delayed CSIT.
\begin{theorem} \label{thm: outer-bound MIMO BC d-CSI}
An outer-bound to the DoF region of the MIMO BC with delayed CSIT is
\begin{eqnarray*}
\mathbf{D}^{\rm{d-CSI}}_{\rm{outer}} = \left\{ \Big( d_i \Big)_{i=1}^K \Big|~  d_i \geq 0 ~ \forall ~ i, ~ \sum_{i=1}^K \frac{d_{\pi(i)}}{\min \left( M, \sum_{j=i}^K N_{\pi(j)} \right) } \leq 1, ~ \forall ~ \pi
\right\} ,
\end{eqnarray*}
where $\pi$ is a permutation of the set $\{1,~ 2,~ \cdots,~ K\}$.
\end{theorem}
\begin{IEEEproof}
We provide two proofs of this theorem. The first one is included in Section \ref{sec: proof of thm: outer-bound MIMO BC d-CSI}, while the second one is presented in Section \ref{sec: proof2 of thm: outer-bound MIMO BC d-CSI}.
\end{IEEEproof}

The above theorem extends the outer-bound of \cite{maddah_ali_tse_delayed_CSIT}, which is applicable to the MISO BC, to the MIMO BC. As mentioned in the introduction, our first proofs uses the ideas of \cite{maddah_ali_tse_delayed_CSIT} and the second proof uses generic techniques in information theory and can be seen as being more widely applicable (cf. \cite{Vaze_Varanasi_delay_MIMO_IC}).

The next theorem proves that the above outer-bound is tight in the two-user case.

\begin{figure} \centering
\includegraphics[bb=130bp 240bp 500bp 550bp,clip,scale=0.75]{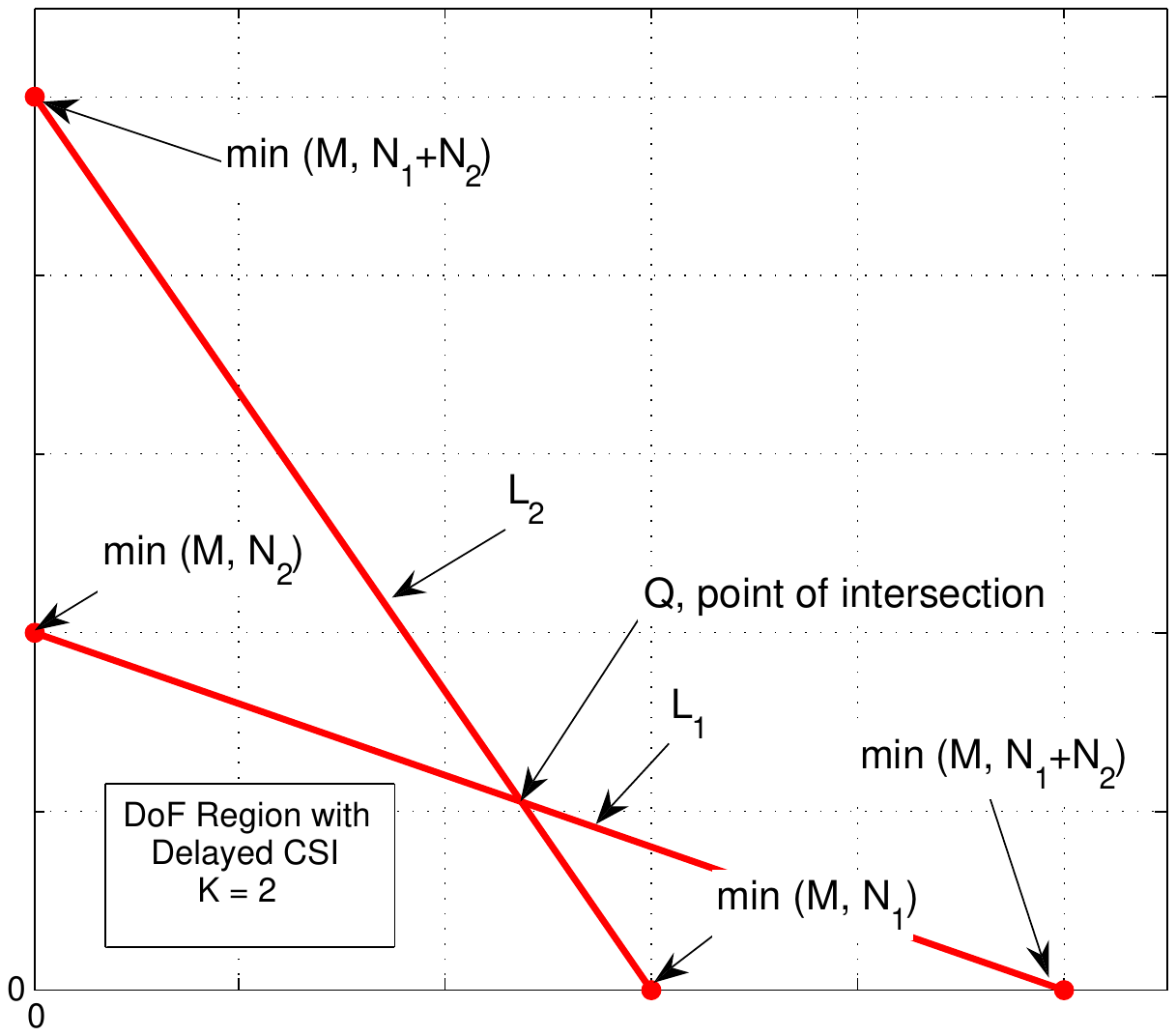}
\caption{The typical shape of the DoF Region of the $2$-user MIMO BC with delayed CSIT} \label{fig: DoF region MIMO BC d-CSI typical shape}
\end{figure}

\begin{theorem} \label{thm: DoF region two-user MIMO BC d-CSI}
For the two-user MIMO BC with delayed CSIT, the outer-bound proposed in Theorem
\ref{thm: outer-bound MIMO BC d-CSI} is achievable. In other words, the DoF region for $K = 2$ is given by
\begin{eqnarray*}
\lefteqn{  \mathbf{D}^{\rm{d-CSI}}_{K = 2} \left(M,N_1,N_2\right) = \Biggl\{ (d_1,d_2) \Big| ~ d_1, d_2 \geq 0,~ \Big. \Biggr. } \\
&& {} \hspace{1cm} \Biggl. \frac{d_1}{\min(M,N_1+N_2)} + \frac{d_2}{\min(M,N_2)} \leq 1, ~ ~ \frac{d_1}{\min(M,N_1)} + \frac{d_2}{\min(M,N_1+N_2)} \leq 1 \Biggr\} .
\end{eqnarray*}
\end{theorem}
\begin{IEEEproof}
See Section \ref{sec: proof of thm: DoF region two-user MIMO BC d-CSI}.
\end{IEEEproof}
The typical shape of the DoF region is shown in Fig. \ref{fig: DoF region MIMO BC d-CSI typical shape}, where $L_1$ and $L_2$ are lines corresponding to the first and second inequality on the weighted sum of $d_1$ and $d_2$, respectively, and $Q$ is the point where they intersect. The intersection of the two triangles formed by $L_1$ and $L_2$ is the DoF region of the 2-user MIMO BC. In the next sub-section, we present the comparison of the DoF regions of the $2$-user MIMO BC under the no CSIT, delayed CSIT, and instantaneous CSIT assumptions.

The following theorem establishes the DoF region of a certain special class of three-user MIMO BCs.

\begin{theorem} \label{thm: DoF region three-user MIMO BC d-CSI}
For the three-user MIMO BC with delayed CSIT and with $N_1 = N_2 = N_3 = N$, and $M \leq 2N$, the outer-bound proposed in Theorem
\ref{thm: outer-bound MIMO BC d-CSI} is achievable. In other words, the delayed-CSIT DoF region for the MIMO BCs with $N_1 = N_2 = N_3 = N$ and $M \leq N$ is given by
\begin{eqnarray*}
\mathbf{D}^{\rm{d-CSI}}_{K = 3} \Big( M ,N, N,N \Big) = \Big\{ (d_1,d_2,d_3) \Big| ~ d_1, d_2,d_3 \geq 0, ~ ~ d_1 + d_2 + d_3 \leq M \Big\} \quad {\rm for} \quad M \leq N,
\end{eqnarray*}
whereas for the MIMO BCs with $N_1 = N_2 = N_3 = N$ and $N < M \leq 2N$, it is given by
\begin{eqnarray*}
\lefteqn{  \mathbf{D}^{\rm{d-CSI}}_{K = 3} \Big( M, N , N , N\Big) = \Big\{ (d_1,d_2) \Big| ~ d_1, d_2 \geq 0,~ \Big. \Big. } \\
&& {} \Big. d_1 + d_2 + \frac{M}{N}d_3 \leq M, ~ d_2 + d_3 + \frac{M}{N}d_1 \leq M, ~ d_3 + d_1 + \frac{M}{N}d_2 \leq M \Big\} \quad {\rm for} \quad N < M \leq 2N .
\end{eqnarray*}
\end{theorem}
\begin{IEEEproof}
See Section \ref{sec: proof of thm: DoF region three-user MIMO BC d-CSI}.
\end{IEEEproof}

\begin{remark}[Comparison of the DoF Regions in the Three-User Case]
Consider a three-user MIMO BC with $N_1 = N_2 = N_3 = N$ and $M \leq 2N$. For such a BC, if $M \leq N$, then the DoF regions with all three assumptions of instantaneous, delayed and no CSIT coincide. On the other hand, when $N < M \leq 2N$, all three regions are (strictly) not equal.
\end{remark}

The following remark provides the delayed-CSIT sum-DoF of a class of $K$-user MIMO BCs using Theorem \ref{thm: outer-bound MIMO BC d-CSI} and the achievable scheme of \cite{maddah_ali_tse_delayed_CSIT}.
\begin{remark}[The Sum-DoF of a Class of MIMO BCs]
Consider the delayed CSIT $K$-user MIMO BC with $N_i = N$ $\geq 1$ $\forall$ $i$ and $M \geq KN$. In \cite{maddah_ali_tse_delayed_CSIT}, it was shown that
\[
d_{sum} \define \max_{(d_1,d_2,\cdots,d_K) \in \mathbf{D}^{\rm{d-CSI}}} \sum_{i=1}^K d_i ~ ~ \geq ~ ~ \frac{KN}{\sum_{i=1}^K \frac{1}{i}}.
\]
Using Theorem \ref{thm: outer-bound MIMO BC d-CSI}, it can be easily shown that $d_{sum} \leq \frac{KN}{\sum_{i=1}^K \frac{1}{i}}$ (cf. \cite{maddah_ali_tse_delayed_CSIT}). Thus, we have the exact characterization for the sum-DoF of this class of MIMO BCs.
\end{remark}

\subsection{A comparison of DoF regions with perfect, delayed, and no CSIT in the two-user case}

\begin{figure} \centering
\includegraphics[bb=100bp 225bp 525bp 560bp,clip,scale=0.75]{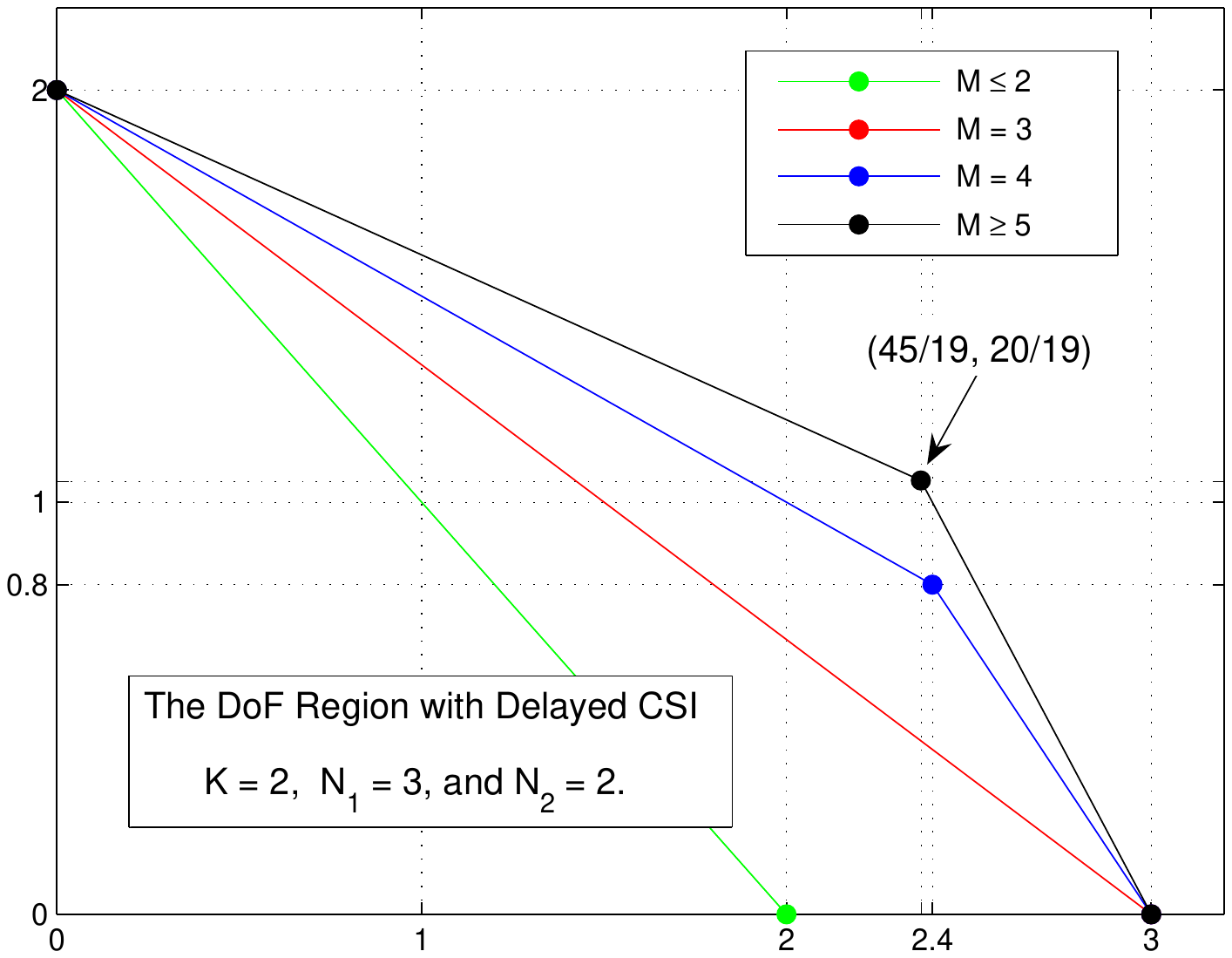}
\caption{The DoF Region of the MIMO BC with $N_1 = 3$ and $N_2 = 2$ for Various Values of $M$} \label{fig: DoF region MIMO BC d-CSI alter with M}
\end{figure}

We first describe the dependence of the DoF region with delayed CSIT on $M$. Consider, for example, the MIMO BC with $N_1 = 3$ and $N_2 = 2$. In Fig. \ref{fig: DoF region MIMO BC d-CSI alter with M}, we show how the DoF region improves with increasing $M$. For small values of $M$, in particular, when $M \leq N_1 = 3$, the DoF region can be achieved without CSIT using time-division. For $M > 3$, interference alignment is needed to achieve the DoF region. As $M$ increases beyond $N_1 + N_2 = 5$, the DoF region remains unchanged.

The DoF region with perfect and global CSIT is known from \cite{Shamai-W-S} and that without CSIT has been derived in \cite{Vaze_Dof_final, Chiachi2}. For small $M$, the DoF region with perfect CSIT can be achieved without CSIT. In particular, if $M \leq N_2$, the DoF region with perfect, delayed, and no CSIT are identical. As $M$ increases beyond $N_2$, the DoF region shrinks in the absence of CSIT. In particular, for $N_2 < M \leq N_1$, the DoF region with perfect CSIT is strictly bigger than the one with delayed CSIT which in turn is equal to that without CSIT. So if $M \leq N_1$, there is no DoF advantage in having delayed CSIT. However, if $M > N_1$, delayed CSIT improves the DoF region; but the region with delayed CSIT is strictly smaller than the one with perfect and instantaneous CSIT.

\section{First Proof of Theorem \ref{thm: outer-bound MIMO BC d-CSI}}
\label{sec: proof of thm: outer-bound MIMO BC d-CSI}

The proof is based on the idea of \cite{maddah_ali_tse_delayed_CSIT}. It is sufficient to prove that
the inequality associated with the identity permutation is a valid outer-bound. First, we outer-bound the capacity region of the given MIMO BC with delayed CSIT (denoted as $\rm{BC}^o$) by assuming that
\begin{enumerate}
\item the $i^{th}$ receiver, in addition to its own outputs, has access to the outputs of receivers $j=i+1 , \cdots , K $; for instance, at each time $t$, it observes $\{Y_j(t)\}_{j=i}^K $
\item at time $t$, the transmitter knows outputs $\{Y_i(n)\}_{i=1}^K$ $\forall$ $n < t$, in addition to delayed CSI.
\end{enumerate}
Clearly, the resulting MIMO BC, call it $\rm{BC}^1$, is physically degraded and its capacity region is an outer-bound to that of $\rm{BC}^o$.

In $\rm{BC}^1$ the side-information available to the transmitter can be considered as being obtained via Shannon-sense feedback from each receiver. However, from the result of \cite{Gamal_fb_capacity_degraded_BC}, feedback can not enhance the capacity region of the physically degraded BC. Hence, the capacity region of $\rm{BC}^1$ remains unchanged even if the transmitter is unaware of $\{Y_i(n), H_i(n)\}_{i=1}^K$ $\forall$ $n<t$ and $\forall$ $t$. Consider now the MIMO BC, denoted as $\rm{BC}^2$, which is identical to $\rm{BC}^1$ except that the transmitter doesn't have this Shannon feedback information. Thus, $\rm{BC}^2$ is a physically-degraded MIMO BC in which there is perfect CSI at the receivers but without CSIT, and in which the $i^{th}$ user has $\sum_{j \geq i} N_j$ receive antennas. Now, the DoF region of $\rm{BC}^2$ can be obtained from the results obtainef by the authors in \cite[Theorem 2]{Vaze_Dof_final}, which yields us the following: if a DoF tuple $(d_1,~ d_2,~ \cdots,~ d_K)$ is achievable over $\rm{BC}^2$, then the inequality $\sum_{i=1}^K \frac{d_i}{\min \left( M, \sum_{j=i}^K N_j \right)} \leq 1$ holds.

Since a DoF-tuple achievable over $\rm{BC}^o$ is also achievable over $\rm{BC}^2$, the above inequality must hold for every DoF-tuple belonging to the DoF region of $\rm{BC}^o$. The rest of the bounds are obtained in the same manner by considering all possible permutations of the user indices. This gives us Theorem \ref{thm: outer-bound MIMO BC d-CSI}.

\section{Second Proof of Theorem \ref{thm: outer-bound MIMO BC d-CSI}} \label{sec: proof2 of thm: outer-bound MIMO BC d-CSI}

As in the first proof, it is sufficient here too to prove the inequality corresponding to the identity permutation. Before starting the main proof, we introduce some notation.

{\em Notation: } The set of all channel matrices at time $t$ is denoted by $H(t)$, i.e., $H(t) = \big\{ H_{i}(t) \big\}_{i=1}^K$. For integers $n_1$ and $n_2$, if $n_1 \leq n_2$, $[n_1:n_2] =  \{n_1,n_1+1, \cdots, n_2\}$; whereas if $n_1 > n_2$, then $[n_1: n_2]$ denotes the empty set. Next, for a $V(t)$ that is a function of $t$ and an $n \geq 1$, $\overline{V}(n) = \big\{ V(t) \big\}_{t=1}^n$. Further, $\overline{Y}_{[n_1:n_2]}(n) = \big\{ \overline{Y}_i(n) \big\}_{i=n_1}^{n_2}$. Similarly, ${\cal M}_{[n_1:n_2]} = \{{\cal M}_i\}_{i=n_1}^{n_2}$ if $n_1,n_2 \in [1:K]$, else it is set equal to $0$. Finally, $o(\log_2 P)$ denotes any real-valued function $x(P)$ of $P$ such that $\lim_{P \to \infty} \frac{x(P)}{\log_2 P} = 0$.

We first outer-bound the capacity region of the given delayed-CSIT MIMO BC by assuming that the $i^{th}$ user knows the channel outputs $Y_j(t)$, $\forall$ $j > i$, instantaneously, and also the messages ${\cal M}_j$, $\forall$ $j > i$. Now, applying Fano's inequality \cite{CT} at the $i^{th}$ user, we obtain
\begin{eqnarray}
R_i & \leq & \frac{1}{n} I\Big( {\cal M}_i; \overline{Y}_{[i:K]}(n), {\cal M}_{[i+1:K]}, \overline{H}(n) \Big) + \epsilon_n \nonumber \\
& = & \frac{1}{n}  I\Big( {\cal M}_i; \overline{Y}_{[i:K]}(n) \Big| {\cal M}_{[i+1:K]}, \overline{H}(n) \Big) + \epsilon_n \label{eq: eq1 bound on Ri dCSI BC journal} \\
& = & \frac{1}{n} h \Big( \overline{Y}_{[i:K]}(n) \Big|  {\cal M}_{[i+1:K]}, \overline{H}(n) \Big) - h \Big( \overline{Y}_{[i:K]}(n) \Big|  {\cal M}_{[i:K]}, \overline{H}(n) \Big) + \epsilon_n, \label{eq: eq2 bound on Ri dCSI BC journal}
\end{eqnarray}
where $\epsilon_n \to 0$ as $n \to \infty$; the equality (\ref{eq: eq1 bound on Ri dCSI BC journal}) follows from the independence of the messages and the channel matrices, whereas the equality (\ref{eq: eq2 bound on Ri dCSI BC journal}) is true because of the definition of the mutual information. Let $\mathsf{N}_j \define \min \left( M, \sum_{k=j}^K N_k \right)$, where $j \in [1:K]$, and $\mathsf{N}_{K+1} = 0$. Then using the bounds on $R_i$ for each $i \in [1:K]$, we obtain
\begin{eqnarray}
\sum_{i=1}^K \frac{R_i}{\mathsf{N}_i} & \leq & \frac{1}{n} \sum_{i=1}^K \frac{1}{\mathsf{N}_i} h \Big( \overline{Y}_{[i:K]}(n) \Big|  {\cal M}_{[i+1:K]}, \overline{H}(n) \Big) \nonumber \\
& & ~ ~ - ~ ~ \frac{1}{n} \sum_{i=1}^K \frac{1}{\mathsf{N}_i}  h \Big( \overline{Y}_{[i:K]}(n) \Big|  {\cal M}_{[i:K]}, \overline{H}(n) \Big) + \epsilon_n \sum_{i=1}^K \frac{1}{\mathsf{N}_i} \nonumber \\
& = &\frac{1}{n} \sum_{i=1}^{K-1} \left\{ \frac{1}{\mathsf{N}_i} h \Big( \overline{Y}_{[i:K]}(n) \Big|  {\cal M}_{[i+1:K]}, \overline{H}(n) \Big) - \frac{1}{\mathsf{N}_{i+1}} h \Big( \overline{Y}_{[i+1:K]}(n) \Big|  {\cal M}_{[i+1:K]}, \overline{H}(n) \Big) \right\} \nonumber \\
& &  ~ ~\! + ~ ~\! \frac{1}{n \cdot \mathsf{N}_K} h \Big( \overline{Y}_K(n) \Big| \overline{H}(n) \Big) - \frac{1}{n \cdot \mathsf{N}_1} h \Big( \overline{Y}_{[1:K]}(n) \Big|  {\cal M}_{[1:K]}, \overline{H}(n) \Big) + \epsilon_n \sum_{i=1}^n \frac{1}{\mathsf{N}_i}. \label{eq: penultimate ineq delayed BC journal}
\end{eqnarray}
after performing simple algebraic manipulations.

We will now bound each term appearing in the above inequality. To bound the argument of the summation over $i$, consider the next lemma.
\begin{lemma} \label{lem: main ineq delayed BC journal}
For a $p \in [1:K-1]$ and $q = p+1$, we have
\begin{eqnarray}
\frac{1}{\mathsf{N}_q} h \Big( \overline{Y}_{[q:K]}(n) \Big|  {\cal M}_{[q:K]}, \overline{H}(n) \Big) \geq \frac{1}{\mathsf{N}_p} h \Big( \overline{Y}_{[p:K]}(n) \Big|  {\cal M}_{[q:K]}, \overline{H}(n) \Big) + n \cdot o(\log_2 P), \label{eq: lem: main ineq delayed BC journal}
\end{eqnarray}
where the term $o(\log_2 P)$ is constant with $n$.
\end{lemma}
\begin{IEEEproof}
See Section \ref{subsec: proof of lem: main ineq delayed BC journal}.
\end{IEEEproof}
We can bound the $i^{th}$ term of the summation in (\ref{eq: penultimate ineq delayed BC journal}) by $\left\{ -n \cdot o(\log_2 P)\right\}$ by applying this lemma with $p = i$.

Consider now the next differential entropy term. Since the DoF of the point-to-point MIMO channel are equal to the minimum of the number of transmit and receive antennas, we have
\begin{equation}
\frac{1}{n} \cdot \frac{1}{\mathsf{N}_K} h \Big( \overline{Y}_K(n) \Big| \overline{H}(n) \Big) = 1 + o(\log_2 P), \label{eq: ptp bound delayed BC journal}
\end{equation}
where $o(\log_2 P)$ is constant with $n$.

Further, since the transmit signal is a deterministic function of the messages, we get
\begin{eqnarray}
\frac{1}{n \cdot \mathsf{N}_1} h \Big( \overline{Y}_{[1:K]}(n) \Big|  {\cal M}_{[1:K]}, \overline{H}(n) \Big) & = & \frac{1}{n \cdot \mathsf{N}_1} h \Big( \overline{Z}_{[1:K]}(n) \Big|  {\cal M}_{[1:K]}, \overline{H}(n) \Big) =  o(\log_2 P). \label{eq: noise term delayed BC journal}
\end{eqnarray}

Thus, the inequalities (\ref{eq: penultimate ineq delayed BC journal}), (\ref{eq: lem: main ineq delayed BC journal}), (\ref{eq: ptp bound delayed BC journal}), and (\ref{eq: noise term delayed BC journal}) yield us
\begin{eqnarray}
\sum_{i=1}^K \frac{R_i}{\mathsf{N}_i}  \leq 1 + o(\log_2 P) + \epsilon_n \sum_{i=1}^n \frac{1}{\mathsf{N}_i}
\end{eqnarray}
on noting that the sum or the difference of the two $o(\log_2 P)$ terms gives another $o(\log_2 P)$ term. Now, first taking the limit as $n \to \infty$ and then as $P \to \infty$, we can obtain the desired inequality.

\begin{remark}[Generalization to the Shannon-feedback case]
This proof can be generalized to prove that the region $\mathbf{D}^{\rm{d-CSI}}_{\rm{outer}}$ is an outer-bound on the DoF region of the MIMO BC with Shannon feedback, where the transmitter knows the channel states and the channel outputs with a finite delay. See also \cite{Vaze_Varanasi_Shannon_fb_2user_IC_journal}, where an outer-bound on the DoF region of the MIMO IC with Shannon feedback is obtained using the techniques that were developed for the MIMO IC with delayed CSIT in \cite{Vaze_Varanasi_delay_MIMO_IC}.
\end{remark}

\begin{remark}[Comparison of the Two Proofs of Theorem \ref{thm: outer-bound MIMO BC d-CSI}]
While the first proof relies on the result of \cite{Gamal_fb_capacity_degraded_BC}, the second proof does not require any such specialized results and makes use of just basic information theoretic identities such as conditioning reduces entropy, chain rule for differential entropy, etc. Hence, the two proofs are fundamentally different.
\end{remark}

\subsection{Proof of Lemma \ref{lem: main ineq delayed BC journal}} \label{subsec: proof of lem: main ineq delayed BC journal}

We use the following notation here.

\begin{figure}
\centering
\includegraphics[scale=0.6]{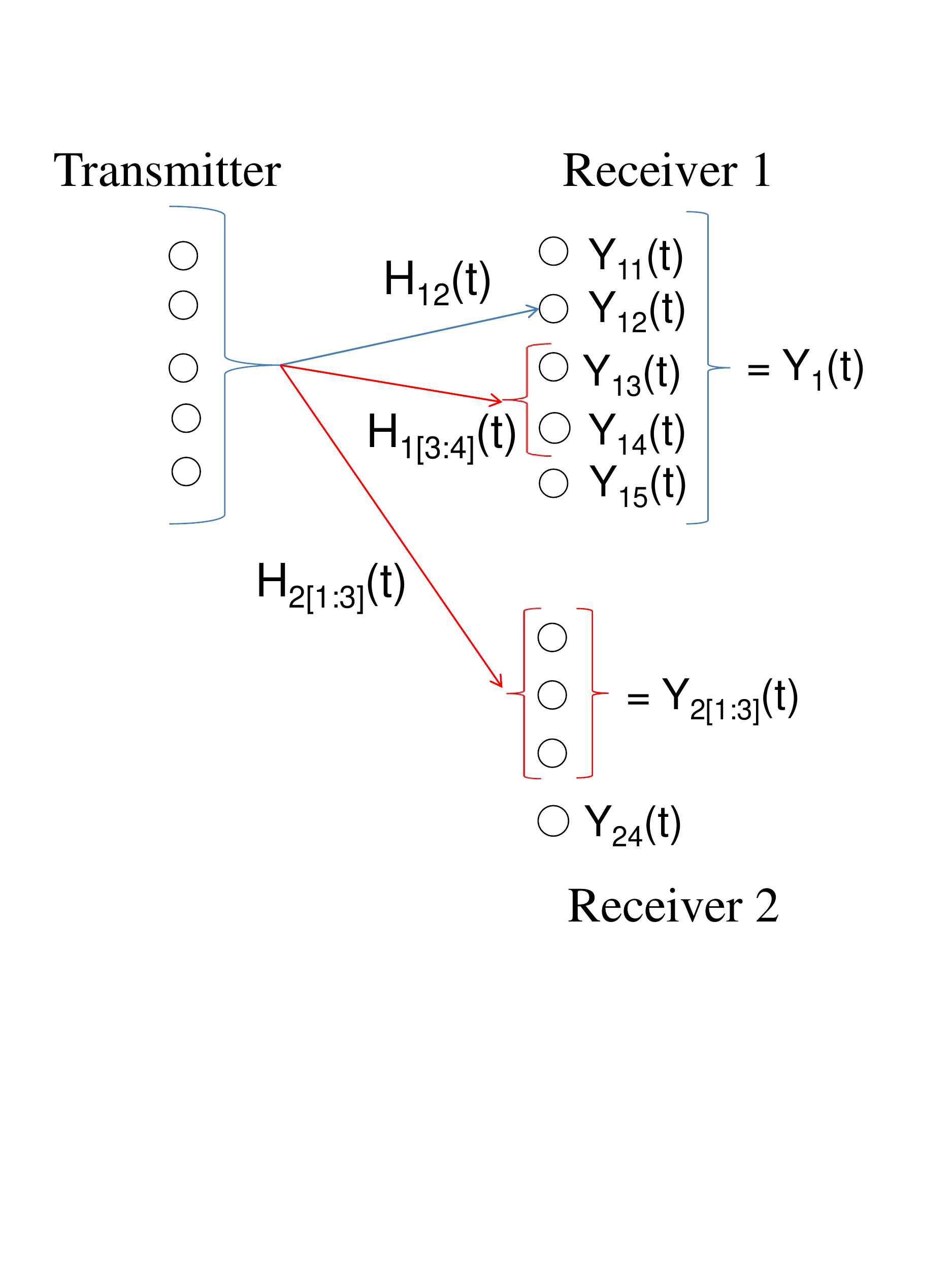}
\caption{Illustrating the Notation Used}
\label{fig: notation used BC journal}
\end{figure}

{\em Notation: } For a random variable $X(t)$, $X([n_1:n_2]) = \{X(t)\}_{t=n_1}^{n_2}$ if $n_1 \leq n_2$, whereas $X([n_1:n_2])$ denotes an empty set if $n_1 > n_2$. For the received signal $Y_i(t)$ and the channel matrix $H_{i}(t)$, the $j^{th}$ entry and the $j^{th}$ row are denoted respectively by $Y_{ij}(t)$ and $H_{ij}(t)$; See Fig. \ref{fig: notation used BC journal}. Further, whenever $n_1 \leq n_2$ and $n_3 \leq n_4$; $Y_{i[n_1:n_2]}(t) = \{Y_{ij}(t)\}_{j=n_1}^{n_2}$, $Y_{i[n_1:n_2]}([n_3:n_4]) = \big\{ \{Y_{ij}(t)\}_{j=n_1}^{n_2} \big\}_{t=n_3}^{n_4}$, $H_{i[n_1:n_2]}(t)$ is the channel matrix from the transmitter to channel outputs $Y_{i[n_1:n_2]}(t)$; however, if $n_1 > n_2$ and/or $n_3>n_4$, then $Y_{i[n_1:n_2]}(t)$ and $Y_{i[n_1:n_2]}([n_3:n_4])$ denote empty sets. Moreover, for $n \geq 1$, $\overline{Y}_{i[n_1:n_2]}(n) = Y_{i[n_1:n_2]}([1:n])$.

This proof is similar to the proof of \cite[Lemma 1]{Vaze_Varanasi_delay_MIMO_IC}. The following two lemmas prove that although the received signal $Y_i(t)$ contains $N_i$ entries, only $\mathsf{N}_i - \mathsf{N}_{i+1}$ of those are relevant ($\mathsf{N}_{K+1} = 0$).
\begin{lemma} \label{lem: derive essential part 1 delayed BC journal}
We have
\begin{eqnarray*}
\lefteqn{  h \Big( \overline{Y}_{[q:K]}(n) \Big|  {\cal M}_{[q:K]}, \overline{H}(n) \Big) } \\
&& {} = h \Big( \overline{Y}_q(n), \overline{Y}_r(n), \cdots, \overline{Y}_K(n) \Big|  {\cal M}_{[q:K]}, \overline{H}(n) \Big) \\
&& {} \geq h \Big( \overline{Y}_{q[1:\mathsf{N}_q-\mathsf{N}_{r}]}(n), \overline{Y}_{r[1:\mathsf{N}_r-\mathsf{N}_{r+1}]}(n), \cdots, \overline{Y}_{K[1:\mathsf{N}_K]}(n)  \Big|  {\cal M}_{[q:K]}, \overline{H}(n) \Big) + n \cdot o(\log_2 P),
\end{eqnarray*}
where $r = q+1$ and the term $o(\log_2 P)$ is constant with $n$.
\end{lemma}
\begin{IEEEproof}
The $q^{th}$ receiver, by our assumption, knows signals $Y_q(t)$, $Y_r(t)$, $\cdots$, $Y_K(t)$ at time $t$. Of the total $\sum_{i=q}^K N_i$ entries of $Y_{[q:K]}(t)$, this receiver can choose any ${\sf N}_q = \min \left( M,\sum_{i=q}^K N_i \right)$ entries, and using them, it can determine the transmit signal $X(t)$ by inverting the channel (since the channel matrices are i.i.d. Rayleigh faded), except for some additive noise term, which is not important in the DoF analysis. Hence, of the total $\sum_{i=q}^K N_i$ entries of $Y_{[q:K]}(t)$, any ${\sf N}_q$ can be retained. In particular, we can choose the first $\mathsf{N}_i - \mathsf{N}_{i+1}$ from $Y_i(t)$, $i \geq q$. This heuristic argument can be proved rigorously using the techniques of \cite[Proof of Lemma 2]{Vaze_Varanasi_delay_MIMO_IC}.
\end{IEEEproof}

\begin{lemma} \label{lem: derive essential part 2 delayed BC journal}
We have
\begin{eqnarray*}
\lefteqn{  h \Big( \overline{Y}_{[p:K]}(n) \Big|  {\cal M}_{[q:K]}, \overline{H}(n) \Big) } \\
&& {} = h \Big( \overline{Y}_p(n), \overline{Y}_q(n), \cdots, \overline{Y}_K(n) \Big|  {\cal M}_{[q:K]}, \overline{H}(n) \Big) \\
&& {} \leq h \Big( \overline{Y}_{p[1:\mathsf{N}_p-\mathsf{N}_{q}]}(n), \overline{Y}_{q[1:\mathsf{N}_q-\mathsf{N}_r]}(n), \cdots, \overline{Y}_{K[1:\mathsf{N}_K]}(n)  \Big|  {\cal M}_{[q:K]}, \overline{H}(n) \Big) + n \cdot o(\log_2 P),
\end{eqnarray*}
where the term $o(\log_2 P)$ is constant with $n$ (recall $q = p+1$).
\end{lemma}
\begin{IEEEproof}
The main idea of this lemma is same as that in Lemma \ref{lem: derive essential part 1 delayed BC journal} so that it can be proved along the lines of the proof of \cite[Lemma 3]{Vaze_Varanasi_delay_MIMO_IC}.
\end{IEEEproof}

To keep the notation simple, we define $W_1(t) \define Y_{p[1:\mathsf{N}_p-\mathsf{N}_{q}]}(t)$
\[
\mbox{and } W_2(t) \define \begin{bmatrix} Y_{q[1:\mathsf{N}_q-\mathsf{N}_{r}]}(t) \\  Y_{r[1:\mathsf{N}_r-\mathsf{N}_{r+1}]}(t) \\ \vdots \\ Y_{K[1:\mathsf{N}_K]}(t) \end{bmatrix} \mbox{ so that } \overline{W}_1(n) = \overline{Y}_{p[1:\mathsf{N}_p-\mathsf{N}_{q}]}(n) \mbox{ and } \overline{W}_2(n) = \begin{bmatrix} \overline{Y}_{q[1:\mathsf{N}_q-\mathsf{N}_{r}]}(n) \\ \overline{Y}_{r[1:\mathsf{N}_r-\mathsf{N}_{r+1}]}(n) \\ \vdots \\ \overline{Y}_{K[1:\mathsf{N}_K]}(n) \end{bmatrix}.
\]
Note that $W_1(t)$ contains ${\sf N}_p - {\sf N}_q$ entries, while $W_2(t)$ contains ${\sf N}_q$ entries.

Consider the following lemma, which is later used in conjunction with Lemmas \ref{lem: derive essential part 1 delayed BC journal} and \ref{lem: derive essential part 2 delayed BC journal} to prove Lemma \ref{lem: main ineq delayed BC journal}.

\begin{lemma}   \label{lem: eq: simplify main ineq delayed BC journal}
The following inequality is true:
\begin{equation}
\frac{1}{\mathsf{N}_q} h \Big( \overline{W}_2(n)  \Big|  {\cal M}_{[q:K]}, \overline{H}(n) \Big) \geq \frac{1}{\mathsf{N}_p}  h \Big( \overline{W}_1(n), \overline{W}_2(n)  \Big|  {\cal M}_{[q:K]}, \overline{H}(n) \Big). \label{eq: simplify main ineq delayed BC journal}
\end{equation}
\end{lemma}
\begin{IEEEproof}
See Section \ref{subsec: proof of lem: eq: simplify main ineq delayed BC journal}.
\end{IEEEproof}

{\em Proof of Lemma \ref{lem: main ineq delayed BC journal}: } Using Lemmas \ref{lem: derive essential part 1 delayed BC journal}, \ref{lem: derive essential part 2 delayed BC journal}, and \ref{lem: eq: simplify main ineq delayed BC journal}, we obtain
\begin{eqnarray*}
\lefteqn{  \frac{1}{\mathsf{N}_q} h \Big( \overline{Y}_{[q:K]}(n) \Big|  {\cal M}_{[q:K]}, \overline{H}(n) \Big) \geq \frac{1}{\mathsf{N}_q} h \Big( \overline{W}_2(n)  \Big|  {\cal M}_{[q:K]}, \overline{H}(n) \Big) + n \cdot o(\log_2 P) }\\
&& {} \geq \frac{1}{\mathsf{N}_p}  h \Big( \overline{W}_1(n), \overline{W}_2(n)  \Big|  {\cal M}_{[q:K]}, \overline{H}(n) \Big) +n \cdot o(\log_2 P) \\
&& {} \geq   \frac{1}{\mathsf{N}_p} h \Big( \overline{Y}_{[p:K]}(n) \Big|  {\cal M}_{[q:K]}, \overline{H}(n) + n \cdot o(\log_2 P),
\end{eqnarray*}
which proves Lemma \ref{lem: main ineq delayed BC journal}.

\subsection{Proof of Lemma \ref{lem: eq: simplify main ineq delayed BC journal}} \label{subsec: proof of lem: eq: simplify main ineq delayed BC journal}

Note first that if ${\sf N}_p - {\sf N}_q = 0$, Lemma \ref{lem: eq: simplify main ineq delayed BC journal} holds trivially. Hence, in the following, we consider the case of ${\sf N}_p - {\sf N}_q > 0$. Let us define $
Q(t) \define \Big\{ \mathcal{M}_{[q:K]}, ~ \overline{H}(t), ~ \overline{W}_2(t-1) \Big\}.$ Consider now the following lemma, which is an important step in the proof of Lemma \ref{lem: eq: simplify main ineq delayed BC journal}.
\begin{lemma}
For an $i \in [1:{\sf N}_q-1]$, and a $k \in [1: {\sf N}_p -{\sf N}_q -1]$, if $j = i+1$ and $l = k+1$,
\begin{eqnarray*}
h \Big( W_{2i}(t) \Big| Q(t),W_{2[1:i-1]}(t) \Big) & = & h \Big( W_{2j}(t) \Big| Q(t),W_{2[1:i-1]}(t) \Big), \\
h \Big( W_{2{\sf N}_q}(t) \Big| Q(t),W_{2[1:{\sf N}_q-1]}(t) \Big) & = & h \Big( W_{11}(t) \Big| Q(t),W_{2[1:{\sf N}_q-1]}(t) \Big), \\
h \Big( W_{1k}(t) \Big| Q(t),W_{2[1:{\sf N}_q]}(t),W_{1[1:k-1]}(t) \Big) & = & h \Big( W_{1l}(t) \Big| Q(t),W_{2[1:{\sf N}_q]}(t),W_{1[1:k-1]}(t) \Big).
\end{eqnarray*}
\label{lem:stateq}
\end{lemma}
\begin{IEEEproof}
This equality follows on noting that conditioned on $Q(t)$ and $W_{2[1:i-1]}(t)$, $W_{2i}(t)$ and $W_{2j}(t)$ are identically distributed.
\end{IEEEproof}

The equalities in the lemma show that the signals received at two different receive antennas of the channel are statistically equivalent when conditioned on the past and the present channel matrices, some of the messages, past received signal, and the present received signal at some other receive antennas of the channel. We refer to this idea as the {\em statistical equivalence of channel outputs}. As we will see, this idea holds the key to this proof of the outer-bound. 

The next lemma is a simple corollary of Lemma \ref{lem:stateq}.
\begin{lemma}
\label{lem:cor}
For an $i \in [1:{\sf N}_q-1]$, and a $k \in [1: {\sf N}_p -{\sf N}_q -1]$, if $j = i+1$ and $l = k+1$,
\begin{eqnarray*}
h \Big( W_{2i}(t) \Big| Q(t),W_{2[1:i-1]}(t) \Big) & \geq & h \Big( W_{2j}(t) \Big| Q(t),W_{2[1:i]}(t) \Big), \\
h \Big( W_{2{\sf N}_q}(t) \Big| Q(t),W_{2[1:{\sf N}_q-1]}(t) \Big) & \geq & h \Big( W_{11}(t) \Big| Q(t),W_{2[1:{\sf N}_q]}(t) \Big), \\
h \Big( W_{1k}(t) \Big| Q(t),W_{2[1:{\sf N}_q]}(t),W_{1[1:k-1]}(t) \Big) & \geq & h \Big( W_{1l}(t) \Big| Q(t),W_{2[1:{\sf N}_q]}(t),W_{1[1:k]}(t) \Big).
\end{eqnarray*}
\end{lemma}
\begin{IEEEproof}
Follows since conditioning reduces entropy.
\end{IEEEproof}

The following lemma can be obtained using Lemma \ref{lem:cor}.
\begin{lemma}
For a $t \in [1:n]$, we have
\[
{\sf N}_p \cdot h \Big( W_2(t) \Big| Q(t) \Big) \geq {\sf N}_q \cdot h \Big( W_1(t), W_2(t) \Big| Q(t), \overline{W}_1(t-1) \Big).
\]
\end{lemma}
\begin{IEEEproof}
Repeated application of the first and the third inequality of the Lemma \ref{lem:cor} yields
\begin{eqnarray*}
h \Big( W_{2i}(t) \Big| Q(t), W_{2[1:i-1]}(t) \Big) & \geq & h \Big( W_{2{\sf N}_q}(t) \Big| Q(t), W_{2[1:{\sf N}_q-1]}(t) \Big) \\
 h \Big( W_{11}(t) \Big| Q(t), W_{2[1:{\sf N}_q]}(t) \Big) & \geq & h \Big( W_{1k}(t) \Big| Q(t),W_{2[1:{\sf N}_q]}(t),W_{1[1:k-1]}(t) \Big)
\end{eqnarray*}
for any $i < {\sf N}_q$ and any $k > 1$. Now, using the chain rule for the differential entropy, we get
\begin{eqnarray}
\frac{1}{{\sf N}_q} h \Big( W_2(t) \Big| Q(t) \Big) & = & \frac{1}{{\sf N}_q} \sum_{i=1}^{{\sf N}_q} h \Big( W_{2i}(t) \Big| Q(t), W_{2[1:i-1]}(t) \Big) \nonumber \\
& \geq &   h \Big( W_{2{\sf N}_q}(t) \Big| Q(t), W_{2[1:{\sf N}_q-1]}(t) \Big) \geq h \Big( W_{11}(t) \Big| Q(t), W_{2[1:{\sf N}_q]}(t) \Big) \label{eq: second ineq applic BC dCSI journal}  \\
& \geq &   \frac{1}{{\sf N}_p - {\sf N}_q} \sum_{k=1}^{{\sf N}_p - {\sf N}_q} h \Big( W_{1k}(t) \Big| Q(t),W_{2}(t),W_{1[1:k-1]}(t) \Big) \nonumber \\
& = &  \frac{1}{{\sf N}_p - {\sf N}_q} h \Big( W_{1}(t) \Big| Q(t), W_{2}(t) \Big), \nonumber
\end{eqnarray}
where the inequality (\ref{eq: second ineq applic BC dCSI journal}) holds due to the second inequality of the previous lemma. Now, since conditioning reduces entropy, we have
\begin{eqnarray}
({\sf N}_p - {\sf N}_q) \cdot h \Big( W_2(t) \Big| Q(t) \Big) & \geq & {\sf N}_q \cdot h \Big( W_{1}(t) \Big| Q(t), W_{2}(t) \Big) \nonumber \\
& \geq & {\sf N}_q \cdot h \Big( W_{1}(t) \Big| Q(t), W_{2}(t), \overline{W}_1(t-1) \Big) \label{eq: eq: simplify main ineq delayed BC journal at t 1}\\
\mbox{and } {\sf N}_q \cdot h \Big( W_2(t) \Big| Q(t) \Big) & \geq & {\sf N}_q  \cdot h \Big( W_2(t) \Big| Q(t), \overline{W}_1(t-1)  \Big). \label{eq: eq: simplify main ineq delayed BC journal at t 2}
\end{eqnarray}
The lemma can now be obtained by adding inequalities (\ref{eq: eq: simplify main ineq delayed BC journal at t 1}) and (\ref{eq: eq: simplify main ineq delayed BC journal at t 2}).
\end{IEEEproof}

{\em Proof of Lemma \ref{lem: eq: simplify main ineq delayed BC journal}: } Using the chain rule for differential entropy,
\begin{eqnarray*}
\lefteqn{ \frac{1}{\mathsf{N}_q} h \Big( \overline{W}_2(n)  \Big|  {\cal M}_{[q:K]}, \overline{H}(n) \Big) = \frac{1}{\mathsf{N}_q} \sum_{t=1}^n  h \Big( W_2(t)  \Big|  {\cal M}_{[q:K]}, \overline{H}(n),\overline{W}_2(t-1) \Big) } \\
&& {} = \frac{1}{\mathsf{N}_q} \sum_{t=1}^n  h \Big( W_2(t)  \Big|  {\cal M}_{[q:K]}, \overline{H}(t),\overline{W}_2(t-1) \Big) \\
&& {} \geq  \frac{1}{\mathsf{N}_p} \sum_{t=1}^n  h \Big( W_1(t), W_2(t)  \Big|  {\cal M}_{[q:K]}, \overline{H}(t),\overline{W}_2(t-1), \overline{W}_1(t-1) \Big) \\
&& {} = \frac{1}{\mathsf{N}_p} h \Big( \overline{W}_1(n), \overline{W}_2(n)  \Big|  {\cal M}_{[q:K]}, \overline{H}(n) \Big),
\end{eqnarray*}
where the second equality holds since the channel matrices $\overline{H}([t+1:n])$ are independent of all other involved random variables; and the subsequent inequality holds due to the last lemma. Hence, Lemma \ref{lem: eq: simplify main ineq delayed BC journal} is proved.


\section{Proof of Theorem \ref{thm: DoF region two-user MIMO BC d-CSI}}
\label{sec: proof of thm: DoF region two-user MIMO BC d-CSI}

The region stated in Theorem \ref{thm: DoF region two-user MIMO BC d-CSI} is shown to be achievable for the two-user MIMO BC with delayed CSIT. From Fig. \ref{fig: DoF region MIMO BC d-CSI typical shape}, it can be seen that it is sufficient to prove that $Q$, the point of intersection of the lines $L_1$ and $L_2$, is achievable because the entire region can then be achieved by time-sharing. The remainder of this section deals with the achievability of the point $Q$.

The analysis is divided into three cases:
\begin{enumerate}
\item Case A: $M \leq N_1$,
\item Case B: $N_1 < M < N_1 + N_2 \Rightarrow N_2 \leq N_1 < M < N_1 + N_2$,
\item Case C: $N_1 + N_2 \leq M \Rightarrow N_2 \leq N_1 < N_1+N_2 \leq M$.
\end{enumerate}

\subsection{Case A: $M \leq N_1$}
In this case, since $\min(M,N_2) < M$, $L_2$ can be easily shown to redundant and then the region defined in Theorem \ref{thm: DoF region two-user MIMO BC d-CSI} is seen to coincide with the DoF region without CSIT \cite{Vaze_Dof_final}. Hence, it is trivially achieved with delayed CSIT.

For the remaining two cases, the DoF region with delayed CSIT is strictly bigger than the one without CSIT. Hence, a transmission scheme to achieve point $Q$ is needed. This scheme happens to be almost identical in the two remaining cases of interest. Therefore, it is described for Case B with an example, and for Case C it is derived in general.

\subsection{Case B: $N_1 < M < N_1 + N_2 \Rightarrow N_2 \leq N_1 < M < N_1 + N_2$}

In this case, the point Q is given by
\begin{eqnarray*}
Q \equiv \left( \frac{M \cdot N_1 \cdot (M-N_2)}{N_1(M-N_2) + M (M-N_1)}, \frac{M \cdot N_2 \cdot (M-N_1)}{N_1(M-N_2) + M ( M-N_1)} \right).
\end{eqnarray*}

Consider an example wherein $M = 4$, $N_1 = 3$, and $N_2 = 2$. Consider the achievability of DoF pair $Q \equiv \Big( \frac{24}{10},~ \frac{8}{10} \Big)$. It will be shown that over $N_1(M-N_2) + M (M-N_1) = 10$ time slots, $M \cdot N_1 \cdot (M-N_2) = 24$ and $M \cdot N_2 \cdot (M-N_1) = 8$ DoF for the two users can be achieved, respectively. Let us divide the duration of $10$ time slots into three phases.

\emph{\underline{Phase One} } consists of $N_1(M-N_2) = 6$ times slots. At each time instant of this phase, the transmitter sends $4$ symbols intended for the first user. Let these data symbols be $\{u_{1i}(j)\}$, where $i \in [1:4]$ and $j \in [1:6]$ \footnote{We adopt the notation that if $n_1$ and $n_2$ are non-negative integers with $n_1 \leq n_2$, then $[n_1 : n_2]$ denotes the set of integers between $n_1$ and $n_2$ (including both).}; and $u_{1i}(j) \sim  \mathcal{C}\mathcal{N} \left(0,\frac{P}{N_1+N_2} \right)$ $\forall$ $i,~j$ and are i.i.d.

Consider the signal received at the first user $\forall ~ t \in [1:6]:$
\[
Y_1(t) = H_1(t) \begin{bmatrix} u_{11}^*(t) & u_{12}^*(t) & u_{13}^*(t) & u_{14}^*(t) \end{bmatrix}^* + Z_1(t).
\]
Thus, for a given $t \in [1:6]$, the first user receives $3$ (noisy) linear combinations of four data symbols $\{u_{1i}(t)\}_{i=1}^4$. Since the channel is taken to be i.i.d. Rayleigh faded, these combinations are linearly independent with probability $1$. This also implies that, for every $t \in [1:6]$, this user needs one more linear combination of $\{u_{1i}(t)\}_{i=1}^4$ so that it can decode the desired symbols.

Even though the second user sees only the interference in this phase, its received signal is still useful as explained below. For a given $t \in [1:6]$, the second user observes two linear combinations of $\{u_{1i}(t)\}_{i=1}^4$, and any one of them is almost surely linearly independent of the three linear combinations seen by the first user. In particular, $\forall ~ t \in [1:6] $, the signal received at the first antenna of the second user is given by
\[
Y_{21}(t) = I_{21}(t) + Z_{21}(t), \mbox{ with } I_{21}(t) = H_{21}(t) \begin{bmatrix} u_{11}^*(t) & u_{12}^*(t) & u_{13}^*(t) & u_{14}^*(t) \end{bmatrix}^* .
\]
Now note that $I_{21}(t)$ is the linear combination of $\{u_{1i}(t)\}_{i=1}^4$ causes interference to the second user but it is useful for the first user. Note that $I_{21}(t)$ is known to the transmitter at the beginning of the $(t+1)^{th}$ time slot due to delayed CSIT. As we will soon see, the transmitter signals over the third phase in such a way that the first receiver learns $I_{21}(t)$ $\forall$ $t \in [1:6]$.

\emph{\underline{Phase 2:} } This phase is analogous to Phase 1 and lasts for  $N_2(M-N_1) = 2$ time slots. In this phase, the transmitter sends $8$ independent data symbols $\{u_{2i}(j)\}$, where $i \in [1: 4]$ and $j \in [1:2]$, to the second user. Its received signal is given for $ ~ \forall ~ t \in [7:8]$ by
\[
Y_2(t) = H_2(t) \begin{bmatrix} u_{21}^*(t') & u_{22}^*(t') & u_{23}^*(t') & u_{24}^*(t') \end{bmatrix}^* + Z_2(t),
\]
where $t' = t-6$. Thus, in order to be able to decode data symbols $\big\{u_{2i}(t-6)\big\}_{i=1}^4$, $t \in [7:8]$, the second user needs two more linear combinations. Moreover, as argued earlier, the two linear combinations are present at any two of the antennas of the first user. The signal received by it at the first two of its antennas over this phase is given for $ t \in [7:8] $ by
\[
\begin{bmatrix} Y_{11}(t) \\ Y_{12}(t) \end{bmatrix} = \begin{bmatrix} I_{11}(t') \\ I_{12}(t') \end{bmatrix} + \begin{bmatrix} Z_{11}(t) \\ Z_{12}(t) \end{bmatrix}
\]
with
\[
 \begin{bmatrix} I_{11}(t') \\ I_{12}(t') \end{bmatrix} = \begin{bmatrix} H_{11}(t) \\ H_{12}(t) \end{bmatrix} \begin{bmatrix} u_{21}^*(t') & \cdots & u_{24}^*(t') \end{bmatrix}^*.
\]
Recall here that $t' = t-6$. In other words, $\{I_{11}(t'),~ I_{12}(t')\}_{t'=1}^2$ are the linear combinations which are useful for the second user.

\emph{\underline{Phase 3:} } The last phase consists of $(M - N_2) (M-N_1) = 2$ time slots. In this phase, the linear combinations $\big\{I_{21}(t)\big\}_{t=1}^6$ are conveyed to the first receiver, while $\big\{ I_{11}(t-6),~ I_{12}(t-6) \big\}_{t=7}^8$ are conveyed to the second. Note that the transmitter knows these linear combinations perfectly at the beginning of Phase 3. Consider the transmit signal for $t = 9,~ 10$:
\begin{eqnarray*}
X(9)   & = & \begin{bmatrix} I_{21}^*(1) & I_{21}^*(2) & I_{11}^*(2) + I_{21}^*(3) & I_{11}^*(1) \end{bmatrix} \mbox{ and }  \\
X(10)  & = & \begin{bmatrix} I_{21}^*(4) & I_{21}^*(5) & I_{12}^*(2) + I_{21}^*(6) & I_{12}^*(1) \end{bmatrix}.
\end{eqnarray*}
Consider time instant $t = 9$. The first user knows\footnote{The first user knows noisy versions of $I_{11}(1)$ and $I_{11}(2)$. But, the presence or absence of noise does not alter the DoF result. It is in this sense that we say that the first user knows $I_{11}(1)$ and $I_{11}(2)$.} $I_{11}(1)$ and $I_{11}(2)$, and thus, can subtract these from the signal it receives at $t = 9$. After removing these linear combinations, it is as if only the first three transmit antennas sent non-zero signals. Hence, the first user can almost surely invert the channel from the first three transmit antennas to its three receive antennas to recover $I_{21}([1:3])$. Similarly, the second user can recover $I_{11}([1,2])$ after subtracting $I_{21}([1:3])$ from its received signal. The operation at time $t = 10$ is similar.

Thus, at the end of $t = 10$, each user receives the required number of linear combinations without any interference. Hence, the DoF tuple under consideration is achievable.

\subsection{Case C: $N_1 + N_2 \leq M \Rightarrow N_2 \leq N_1 < N_1+N_2 \leq M$}

Point Q in this case is given by
\[
Q \equiv \left( \frac{N_1^2 \cdot (N_1 + N_2)}{N_1^2 + N_2^2 + N_1 N_2},~ \frac{N_2^2 \cdot (N_1 + N_2)}{N_1^2 + N_2^2 + N_1 N_2} \right).
\]

The achievability scheme in general consists of three phases as described in the previous section. Moreover, it is sufficient to use only $N_1 + N_2$ transmit antennas. Hence, in the remainder of this subsection, we assume without loss of generality that $M = N_1 + N_2$.

\emph{\underline{Phase 1} } consists of $N_1^2$ time slots. At each time instant, the transmitter sends one independent data symbol intended for the first user per antenna. Thus, a total of $N_1^2 (N_1 + N_2)$ symbols are sent. Let the symbols be $\{u_{1i}(j)\}$, where $i \in [1 : N_1+N_2]$ and $j \in [1: N_1^2]$.
At time $t \in [1: N_1^2]$, the first user gets $N_1$ distinct linear combinations of $\big\{ u_{1i}(t) \big\}_{i=1}^{N_1+N_2}$, and the $N_2$ linear combinations observed by the second user, which are denoted as $\big\{ I_{2j}(t) \big\}_{j=1}^{N_2}$, are useful for the first user.

\emph{\underline{Phase 2} } lasts for the next $N_2^2$ time slots. The transmitter sends independent data symbols $\{u_{2i}(j)\}$, where $i \in [1 : N_1+N_2]$ and $j \in [1: N_2^2]$, intended for the second user. At time $t \in [N_1^1 + 1: N_1^2 + N_2^2]$, the second user receives $N_2$ linear combinations of $\big\{ u_{2i}(t-N_1^2) \big\}_{i=1}^{N_1+N_2}$ and the $N_1$ more linear combinations needed for the second user are observed by the first user as interferences $\big\{ I_{1i}(t-N_1^2) \big\}_{i=1}^{N_1}$ for each $t \in [N_1^1 + 1: N_1^2 + N_2^2]$.

\emph{\underline{Phase 3} } takes the remaining $N_1 N_2$ time slots. The first user has to learn $N_1^2 N_2$ linear combinations $\{I_{2i}(t)\}$, where $i \in [1 : N_2]$ and $t \in [1: N_1^2]$; whereas the second receiver requires $N_2^2 N_1$ linear combinations $\big\{ I_{1i}(t-N_1^2) \big\}$, where $i \in [1: N_1]$ and $t \in [N_1^1 + 1: N_1^2 + N_2^2]$. Moreover, these linear combinations are known to the transmitter at the beginning of Phase 3.

First, partition the set of $N_1^2 N_2$ linear combinations $\big\{ I_{2i}(t) \big\}_{i,t}$ into $N_1 N_2$ disjoint subsets each of cardinality $N_1$. After partitioning, denote these linear combinations as $\big\{ I^{[2]}_{j}(k) \big\}$, where $j \in [1: N_1]$ and $k \in [1 : N_1 N_2]$. Similarly, partition the set $\big\{ I_{1i}(t-N_1^2) \big\}_{i,t}$ into $N_1 N_2$ disjoint subsets of cardinality $N_2$ each; and after partitioning denote these by $\big\{ I^{[1]}_{j}(k) \big\}$ for $j \in [1: N_2]$ and $k \in [1 : N_1 N_2]$. This procedure of partitioning the linear combinations is deterministic and is known to all terminals.

Then at any time $t \in [N_1^2 + N_2^2 + 1 : N_1^2 + N_2^2 + N_1 N_2]$, form the transmit signal as follows:
\[
X(t) = \begin{bmatrix} I^{[2]}_1(t') \\ I^{[2]}_2(t') \\ \vdots \\ I^{[2]}_{N_1}(t') \\ 0_{N_2} \end{bmatrix} + \begin{bmatrix} 0_{N_1} \\ I^{[1]}_1(t') \\ I^{[1]}_2(t') \\ \vdots \\ I^{[1]}_{N_2}(t') \end{bmatrix},
\]
where $t' = t - N_1^2 - N_2^2$ and $0_x$ denotes the column vector consisting of all zeros of length $x$. Since the first user knows $\big\{ I^{[1]}_{j}(t') \big\}_j$, it can subtract these to recover $\big\{ I^{[2]}_{j}(t') \big\}$ via channel inversion. Similarly, the second user can recover $\big\{ I^{[1]}_{j}(t') \big\}_j$.

At the end of the third phase, each user gets the required number of linear combinations without interference and thus can recover its data symbols.

\section{Proof of Theorem \ref{thm: DoF region three-user MIMO BC d-CSI}: $N < M \leq 2N$} \label{sec: proof of thm: DoF region three-user MIMO BC d-CSI}

Note that the converse argument follows from Theorem \ref{thm: outer-bound MIMO BC d-CSI}. Hence, we focus below on the achievability part. Consider first the case where $M \leq N$. Here, the region $\mathbf{D}^{\rm{d-CSI}}_{K = 3} \Big( M, N, N,N \Big)$ with $M \leq N$ is achievable even without CSIT \cite{Vaze_Dof_final}, and hance, also with delayed CSIT. Thus, it is sufficient to deal with the case of $N < M < 2N$, which is the topic of the remainder of this section. We prove that the region $\mathbf{D}^{\rm{d-CSI}}_{K = 3} \Big( M, N, N,N \Big)$ is achievable when $N < M \leq 2N$. Throughout the rest of this section, the inequality $N < M \leq 2N$ holds.

Let us first consider the following lemma, which allows us to express the region $\mathbf{D}^{\rm{d-CSI}}_{K = 3} ( M, N , N , N)$ as the union of the three regions, and thereby simplifies the proof of the theorem.
\begin{lemma}
Suppose, for an $i \in \{1,2,3\}$,
\[
\mathbf{D}_i \define \Big\{ (d_1,d_2,d_3) \in \mathbf{D}^{\rm{d-CSI}}_{K = 3} \Big( M, N , N , N \Big)\Big| ~ N < M \leq 2N, ~ d_i \leq \frac{MN}{M+N} \Big. \Big\}.
\]
Then
\[
\mathbf{D}^{\rm{d-CSI}}_{K = 3} \Big( M, N , N , N \Big) = \bigcup_{i=1}^3 \mathbf{D}_i,
\]
if $N < M \leq 2N$.
\end{lemma}
\begin{IEEEproof}
A 3-tuple $(d_1,d_2,d_3)$ with $d_1$, $d_2$, $d_3$ $> \frac{MN}{M+N}$ can not belong to the region $\mathbf{D}^{\rm{d-CSI}}_{K = 3} \Big( N < M \leq 2N, N , N , N \Big)$ because none of the three bounds on the weighted sums of $d_1$, $d_2$, and $d_3$ present in the definition of $\mathbf{D}^{\rm{d-CSI}}_{K = 3} \Big( N < M \leq 2N, N , N , N \Big)$ would get satisfied at such a 3-tuple. Hence, at least one of $d_1$, $d_2$, and $d_3$ must be less than or equal to $\frac{MN}{M+N}$.
\end{IEEEproof}
Thus, due to this lemma and symmetry, it is sufficient to prove that the region $\mathbf{D}_3$ is achievable, which is the goal henceforth.

We introduce some notation.
\begin{eqnarray*}
{\cal S}(x) & \define & \Big\{ (d_1,d_2,d_3) \in \mathbf{D}^{\rm{d-CSI}}_{K = 3} \Big( M, N , N , N \Big) \Big| ~ N < M \leq 2N, ~ \mbox{ and } d_3 = x \Big. \Big\}, \\
{\cal P}_{ij} \Big\{ (d_1,d_2,d_3) \Big\} & \define & (d_i,d_j), ~i,j\in\{1,2,3\}, \mbox{ and } \\
{\cal P}_{ij} \Big\{ {\cal S}  \Big\} & \define & \Big\{ (d_i,d_j) \Big| ~ (d_1,d_2,d_3) \in {\cal S} \Big\}.
\end{eqnarray*}
Here, ${\cal S}(x)$ is the plane corresponding to $d_3 = x$, while ${\cal P}_{ij}$ represents a projection operation.

From the definition of $\mathbf{D}^{\rm{d-CSI}}_{K = 3} \Big( N < M \leq 2N, N , N , N \Big)$, we observe that the plane ${\cal S}(d_3)$ is defined in terms of the following three bounds:
\begin{eqnarray*}
\begin{array}{ccccc}
L_0(d_3) & \define & d_1 + d_2 & \leq & M - \frac{M}{N} d_3 \\
L_1(d_3) & \define & \frac{M}{N} d_1 + d_2 & \leq &  M - d_3 \\
L_2(d_3) & \define & d_1 + \frac{M}{N} d_2 & \leq & M - d_3.
\end{array}
\end{eqnarray*}
Suppose $P_{ij}(d_3)$ denotes the point of intersection of lines corresponding to $L_i(d_3)$ and $L_j(d_3)$; whereas $P_{id_k}(d_3)$ represents the point at which the line corresponding to $L_i(d_3)$ and $d_k$-axis intersect.

Our goal now is to show that for any $d_3 \in \left[ 0, \frac{MN}{M+N} \right]$, the plane ${\cal S}(d_3)$ is achievable. We divide the proof into five parts corresponding to
\[
d_3 = \frac{MN}{M+N}, ~ d_3 = \frac{MN}{M+2N}, ~ d_3 = 0, ~ d_3 \in \left( \frac{MN}{M+2N}, \frac{MN}{M+N} \right) ~ \mbox{ and } ~ d_3 \in \left( 0, \frac{MN}{M+2N} \right),
\]

\begin{figure}
\includegraphics[scale=0.6]{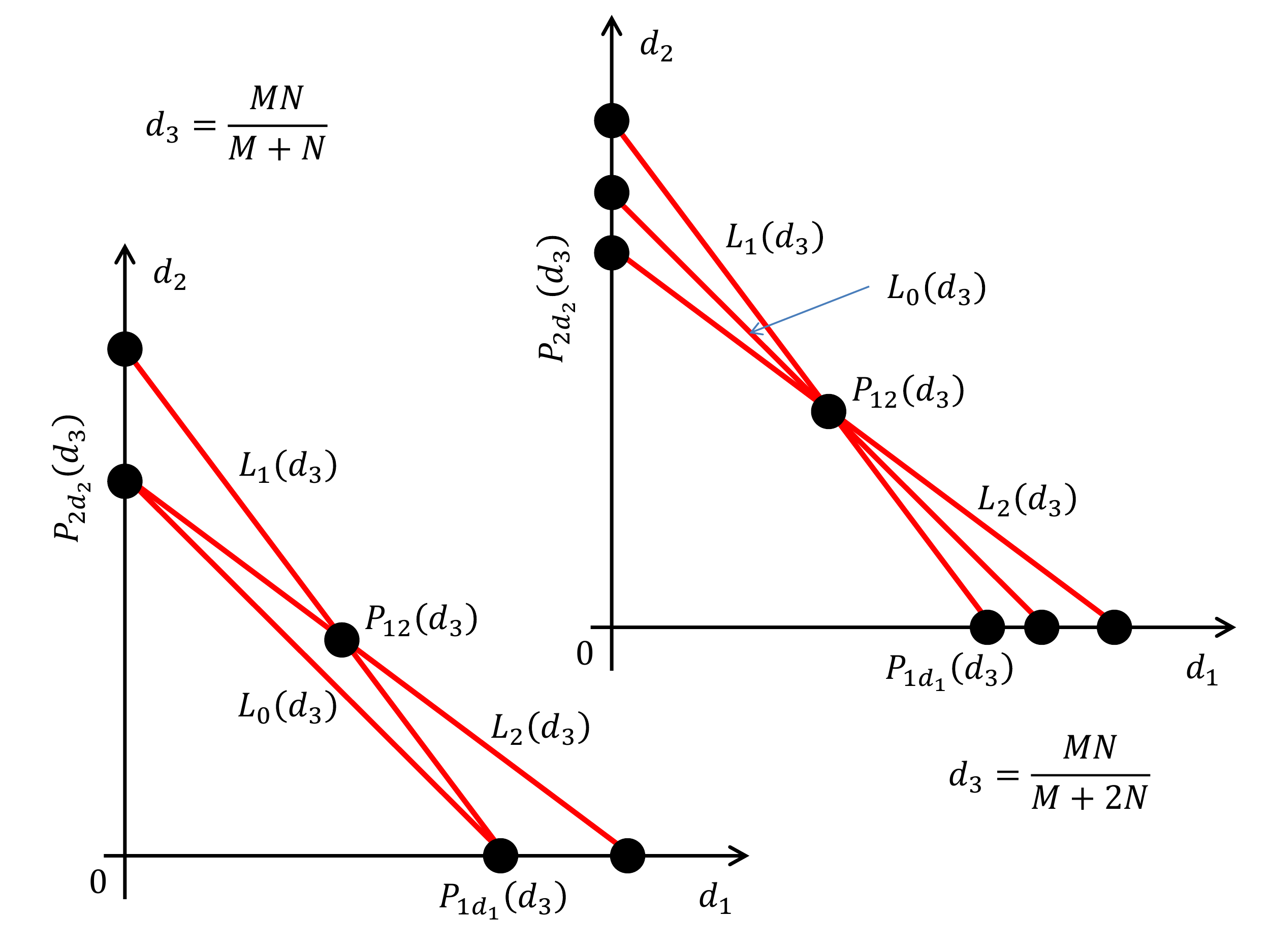}
\caption{Shape of the Plane ${\cal S}(d_3)$ for $d_3 = \frac{MN}{M+N}$ and $d_3 = \frac{MN}{M+2N}$}
\label{fig: plane S d_3 =d_3maxmid journal delayed CSIT MIMO BC}
\end{figure}

\underline{1) $d_3 = \frac{MN}{M+N}$ : } The shape of the plane ${\cal S}\left( \frac{MN}{M+N} \right)$ has been shown in Fig. \ref{fig: plane S d_3 =d_3maxmid journal delayed CSIT MIMO BC}, from which bounds $L_1(d_3)$ and $L_2(d_3)$ are redundant at $d_3 = \frac{MN}{M+N}$. Thus, for $ d_3 = \frac{MN}{M+N}$, it is sufficient to prove the achievability of points
\begin{eqnarray*}
P_{01}(d_3) & \equiv & P_{1d_1}(d_3) \equiv P_{0d_1}(d_3) \equiv \left( \frac{MN}{M+N}, 0, \frac{MN}{M+N} \right),  \\
P_{02}(d_3) & \equiv & P_{2d_2}(d_3) \equiv P_{0d_2}(d_3) \equiv \left( 0, \frac{MN}{M+N}, \frac{MN}{M+N} \right)
\end{eqnarray*}
Consider now the point $P_{1d_1}(d_3)$. Note that
\[
{\cal P}_{13} \Big( P_{1d_1}(d_3) \Big) = \left( \frac{MN}{M+N}, \frac{MN}{M+N} \right) \in \mathbf{D}^{\rm{d-CSI}}_{K = 2} \Big( M , N , N \Big) \quad {\rm for} \quad d_3 = \frac{MN}{M+N}.
\]
Hence, point $P_{01}(d_3)$ is achievable by not transmitting to the $2^{nd}$ user at all and by using the scheme of the last section for the first and the third user. Similarly, the point $P_{2d_2}(d_3)$ can be achieved. Hence, the plane ${\cal S}\left( \frac{MN}{M+N} \right)$ is achievable.

\underline{ 2) $d_3 = \frac{MN}{M+2N}$ : } The shape of the plane ${\cal S}\left( \frac{MN}{M+2N} \right)$ is shown in Fig. \ref{fig: plane S d_3 =d_3maxmid journal delayed CSIT MIMO BC}, from which we observe that the bound $L_0(d_3)$ is redundant and it passes through point $P_{12}(d_3)$. Here, we have
\begin{eqnarray*}
P_{1d_1}(d_3) & \equiv & \left( \frac{N}{M}(M - d_3), 0, d_3 \right) \quad {\rm for} \quad d_3 = \frac{MN}{M+2N} \\
P_{2d_2}(d_3) & \equiv & \left(0, \frac{N}{M}(M-d_3), d_3 \right), \\
P_{12}(d_3) \equiv P_{01}(d_3) \equiv P_{02}(d_3) & \equiv & \left( \frac{MN}{M+2N}, \frac{MN}{M+2N}, \frac{MN}{M+2N} \right).
\end{eqnarray*}
Again, note that the achievability of points $P_{1d_1}(\frac{MN}{M+2N}) \mbox{ and } P_{2d_2}(\frac{MN}{M+2N})$ can be proved as done before in the case of $d_3 = \frac{MN}{M+N}$. Moreover, the achievability of the point
\[
P_{12} \left( \frac{MN}{M+2N} \right)
\]
has been proved in \cite{abdoli_3user_BC_delayed_isit}. Hence, the plane is achievable.

\begin{figure}
\includegraphics[scale=0.7]{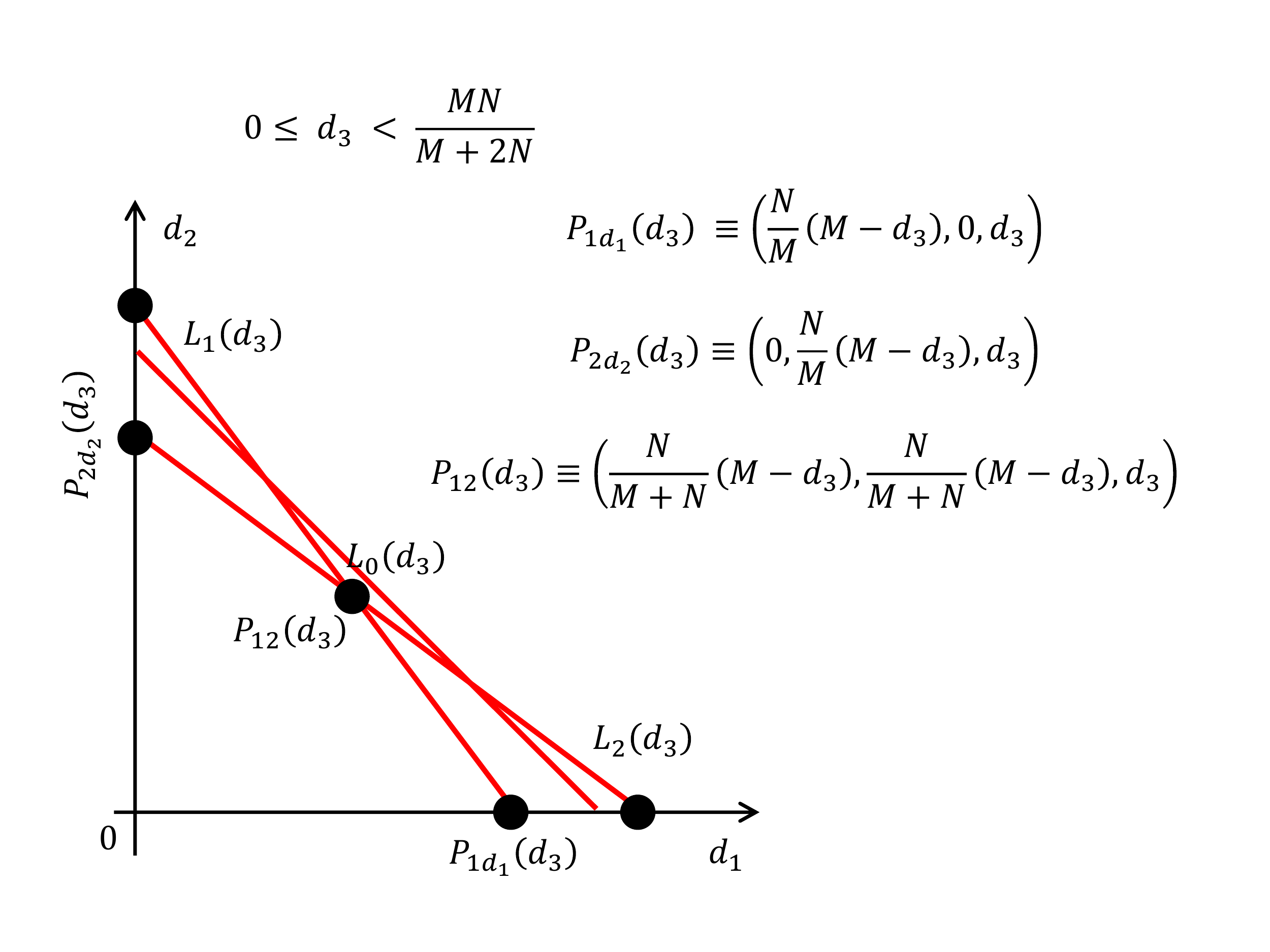}
\caption{Shape of the Plane ${\cal S}(d_3)$ for $0 \leq d_3 \leq \frac{MN}{M+2N}$}
\label{fig: plane S d_3 etn 0 and mid journal delayed CSIT MIMO BC}
\end{figure}

\underline{3) $d_3 = 0$ : } The shape of ${\cal S}\left( 0 \right)$ has been shown in Fig. \ref{fig: plane S d_3 etn 0 and mid journal delayed CSIT MIMO BC}. Bound $L_0(0)$ is redundant and
\[
{\cal P}_{12} \Big\{ {\cal S}(0) \Big\}  =  \mathbf{D}^{\rm{d-CSI}}_{K = 2} \Big( M , N , N \Big).
\]
Hence, the plane ${\cal S}(0)$ is achievable by turning off the third user.

\begin{figure}
\includegraphics[scale=0.7]{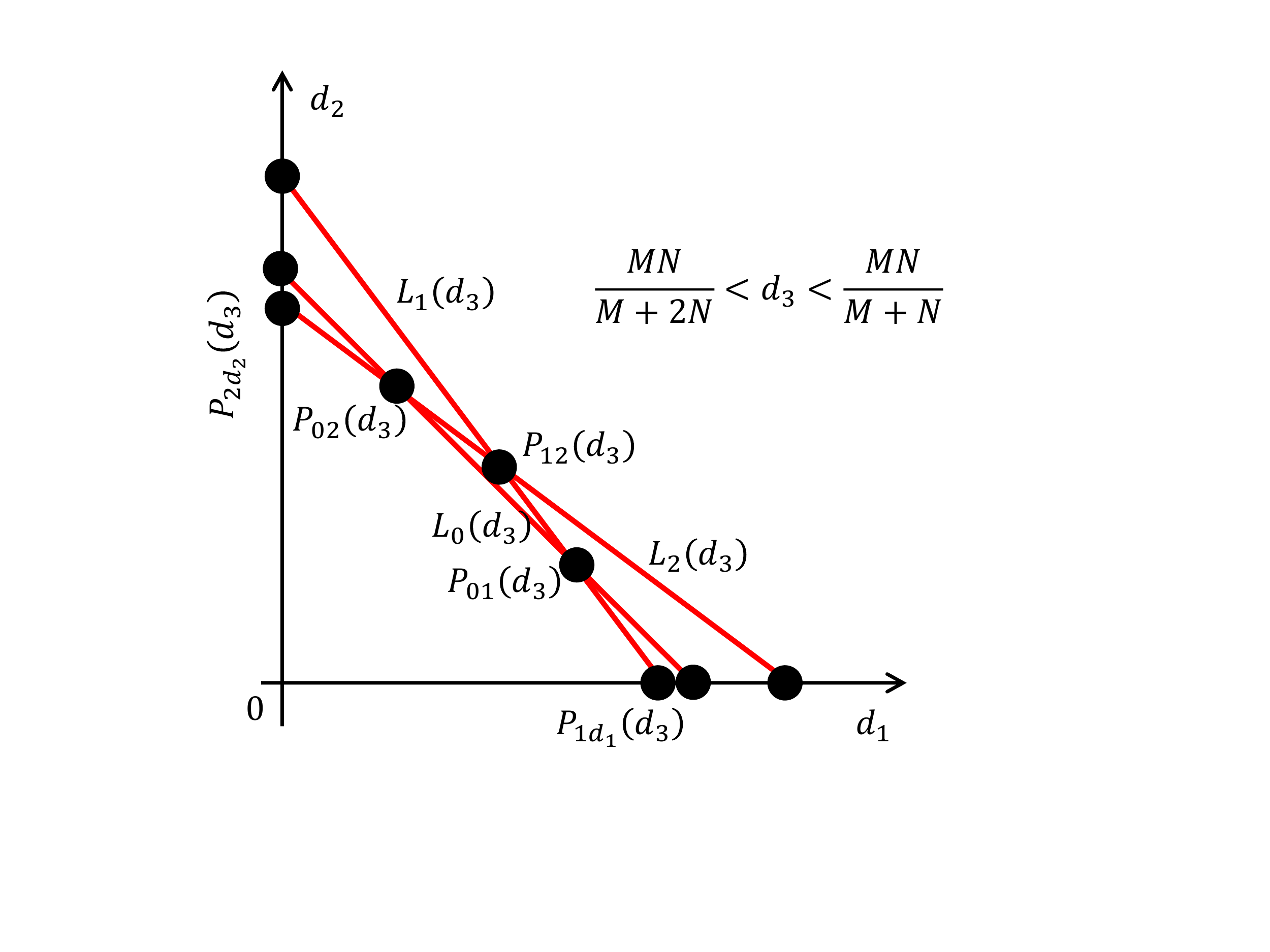}
\caption{Shape of the Plane ${\cal S}(d_3)$ for $\frac{MN}{M+2N} < d_3 < \frac{MN}{M+N}$}
\label{fig: plane S d_3 betn max and mid journal delayed CSIT MIMO BC}
\end{figure}

\underline{4) $ \frac{MN}{M+N} > d_3 > \frac{MN}{M+2N} $ : } The general shape of the plane ${\cal S}(d_3)$ is shown in Fig. \ref{fig: plane S d_3 betn max and mid journal delayed CSIT MIMO BC}. By symmetry, it is sufficient to consider points $P_{01}(d_3)$ and $P_{1d_1}(d_3)$. We can obtain
\begin{eqnarray*}
P_{01}(d_3) & \equiv & \left( d_3, M - \left[ 1+\frac{M}{N} \right] d_3, d_3 \right) \mbox{ and }\\
P_{1d_1}(d_3) & \equiv & \left( \frac{N}{M} \left[ M-d_3 \right], 0 , d_3 \right)
\end{eqnarray*}
after some simple calculations. Furthermore, the point $P_{1d_1}(d_3)$ can be achieved by not transmitting to the second user and by using the scheme of the last section for the remaining two users. It can be verified that the point $P_{01}(d_3)$ can be achieved via time sharing of
\[
P_{01} \Big( \frac{MN}{M+N} \Big) \mbox{ and } P_{01} \Big( \frac{MN}{M+2N} \Big).
\]

\underline{5) $\frac{MN}{M+2N} > d_3 > 0$ : } The general shape of the plane ${\cal S}(d_3)$ is shown in Fig. \ref{fig: plane S d_3 etn 0 and mid journal delayed CSIT MIMO BC}. Again, it is sufficient to focus just on the point $P_{12}(d_3)$, and it can be achieved via time sharing of
\[
P_{12} \Big( \frac{MN}{M+2N} \Big) \mbox{ and } P_{12}\Big( 0 \Big).
\]
Hence, the plane ${\cal S}(d_3)$ is achievable.

This completes the proof of the achievability of $\mathbf{D}_3$. The theorem is hence proved.

\section{Conclusion}
For the $K$-user MIMO BC with delayed CSIT, an outer-bound to its DoF region is obtained. An interference alignment scheme is specified for the two-user MIMO BC that achieves this outer-bound, thereby characterizing the DoF region in this case. For the three-user MIMO BC, the DoF region is characterized for the special class in which there are an equal number of antennas at the receivers and the number of transmit antennas are no more than twice the number of antennas at each of the receivers.

\bibliographystyle{IEEEtran}
\bibliography{v2_delayed_CSIT_MIMO_BC_arxiv}

\begin{thebibliography}{10}
\providecommand{\url}[1]{#1}
\csname url@samestyle\endcsname
\providecommand{\newblock}{\relax}
\providecommand{\bibinfo}[2]{#2}
\providecommand{\BIBentrySTDinterwordspacing}{\spaceskip=0pt\relax}
\providecommand{\BIBentryALTinterwordstretchfactor}{4}
\providecommand{\BIBentryALTinterwordspacing}{\spaceskip=\fontdimen2\font plus
\BIBentryALTinterwordstretchfactor\fontdimen3\font minus
  \fontdimen4\font\relax}
\providecommand{\BIBforeignlanguage}[2]{{%
\expandafter\ifx\csname l@#1\endcsname\relax
\typeout{** WARNING: IEEEtran.bst: No hyphenation pattern has been}%
\typeout{** loaded for the language `#1'. Using the pattern for}%
\typeout{** the default language instead.}%
\else
\language=\csname l@#1\endcsname
\fi
#2}}
\providecommand{\BIBdecl}{\relax}
\BIBdecl

\bibitem{Shamai-W-S}
H.~Weingarten, Y.~Steinberg, and S.~Shamai, ``The capacity region of
  multiple-input multiple-output broadcast channels,'' \emph{IEEE Trans.
  Inform. Theory}, vol. 52, no. 9, pp. 3936--3964, Sep. 2006.

\bibitem{Vaze_Dof_final}
C.~S. Vaze and M.~K. Varanasi, ``The degrees of freedom regions of {M}{I}{M}{O}
  broadcast, interference, and cognitive radio channels with no {C}{S}{I}{T},''
  Sep. 2009, Available Online: http://arxiv.org/abs/0909.5424.

\bibitem{Jafar-Goldsmith}
S.~A. Jafar and A.~J. Goldsmith, ``Isotropic fading vector broadcast channels:
  The scalar upper bound and loss in degrees of freedom,'' \emph{IEEE Trans.
  Inform. Theory}, vol. 51, no. 3, pp. 848--857, Mar. 2005.

\bibitem{Chiachi2}
C.~Huang, S.~A. Jafar, S.~Shamai, and S.~Vishwanath, ``On degrees of freedom
  region of {M}{I}{M}{O} networks without {C}{S}{I}{T},'' Sep. 2009, Available
  Online: http://arxiv.org/pdf/0909.4017.

\bibitem{Jindal}
N.~Jindal, ``M{I}{M}{O} broadcast channels with finite rate feedback,''
  \emph{IEEE Trans. Inform. Theory}, vol. 52, no. 11, pp. 5045--5060, Nov.
  2006.

\bibitem{Ravindran}
N.~Ravindran and N.~Jindal, ``Limited feedback-based block diagonalization for
  the {M}{I}{M}{O} broadcast channel,'' \emph{IEEE Journal on Sel. Areas in
  Comm.}, vol. 26, no. 8, pp. 1473--1482, Oct. 2008.

\bibitem{Caire-Jindal}
G.~Caire, N.~Jindal, M.~Kobayashi, and N.~Ravindran, ``Multiuser {M}{I}{M}{O}
  downlink made practical: achievable rates and simple channel state estimation
  and feedback schemes,'' \emph{submitted to IEEE Trans. Inform. Theory}, Nov.
  2007.

\bibitem{Vaze_fb_scaling_GBC}
C.~S. Vaze and M.~K. Varanasi, ``{C}{S}{I} feedback scaling rate vs
  multiplexing gain tradeoff for {D}{P}{C}-based transmission in the {G}aussian
  {M}{I}{M}{O} broadcast channel,'' in \emph{IEEE Inter. Symp. Inform. Theory},
  Jun. 2010.

\bibitem{maddah_ali_tse_delayed_CSIT}
M.~A. Maddah-Ali and D.~Tse, ``Completely stale transmitter channel state
  information is still very useful,'' Oct. 2010, Available:
  http://arxiv.org/abs/1010.1499.

\bibitem{Gamal_fb_capacity_degraded_BC}
A.~E. Gamal, ``The feedback capacity of degraded broadcast channels,''
  \emph{IEEE Trans. Inform. Theory}, vol. 24, no. 3, pp. 379--381, Apr. 1978.

\bibitem{Vaze_Varanasi_delay_MIMO_IC}
C.~S. Vaze and M.~K. Varanasi, ``The degrees of freedom region and interference
  alignment for the {M}{I}{M}{O} interference channel with delayed {C}{S}{I},''
  \emph{submitted to IEEE Trans. Inform. Th.}, Jan. 2011, Available:
  http://arxiv.org/abs/1101.5809.

\bibitem{abdoli_3user_BC_delayed_isit}
M.~J. Abdoli, A.~Ghasemi, and A.~K. Khandani, ``On the degrees of freedom of
  three-user {M}{I}{M}{O} broadcast channel with delayed {C}{S}{I}{T},'' in
  \emph{IEEE Intern. Symp. Inform. Th.}, St. Petersburg, Russia, Aug. 2011.

\bibitem{Jafar_Shamai_retrospective_IA}
H.~Maleki, S.~A. Jafar, and S.~Shamai, ``Retrospective interference
  alignment,'' Sep. 2010, Available:
  http://arxiv.org/PS\_cache/arxiv/pdf/1009/1009.3593v1.pdf.

\bibitem{Ghasemi_Motahari_Khandani_MIMO_Delayed}
\BIBentryALTinterwordspacing
A.~Ghasemi, A.~Motahari, and A.~Khandani, ``Interference alignment for the mimo
  interference channel with delayed local csit,'' \emph{CoRR}, 2011. [Online].
  Available: \url{http://arxiv.org/abs/1102.5673}
\BIBentrySTDinterwordspacing

\bibitem{Ghasemi_Motahari_Khandani_X_Delayed}
------, ``On the degrees of freedom of x channel with delayed csit,'' in
  \emph{Proceedings of IEEE International Symposium on Information Theory
  (ISIT)}, 2011.

\bibitem{Jafar_correlations}
S.~A. Jafar, ``Exploiting channel correlations - simple interference alignment
  schemes with no {C}{S}{I}{T},'' Oct. 2009, Available Online:
  http://arxiv.org/pdf/0910.0555.

\bibitem{Jafar_Gou_blind_IA_2010}
C.~Wang, T.~Gou, and S.~A. Jafar, ``Aiming perfectly in the dark - blind
  interference alignment through staggered antenna switching,'' Feb. 2010,
  Available Online: http://arxiv.org/abs/1002.2720.

\bibitem{CT}
T.~Cover and J.~Thomas, \emph{Elements of Inform. Theory}.\hskip 1em plus 0.5em
  minus 0.4em\relax John Wiley and Sons, Inc., 1991.

\bibitem{Vaze_Varanasi_Shannon_fb_2user_IC_journal}
C.~S. Vaze and M.~K. Varanasi, ``The degrees of freedom region of the
  {M}{I}{M}{O} interference channel with {S}hannon feedback,'' 2011, under
  preparation.

\end{thebibliography}
\end{document}